\documentclass[a4paper,12pt]{article}

\usepackage{amsmath,amssymb,epsfig,eucal,cite}

\newcommand{\aaa}{\mbox{\sl a}}
\newcommand{\as}{\alpha_\mathrm{s}}      
\newcommand{\asb}{\bar{\alpha}_\mathrm{s}}             

\newcommand{\C}{\mathbb{C}}              
\newcommand{\contour}{\CMcal{C}}        
\newcommand{\dif}{\mathrm{d}}            
\newcommand{\e}{\varepsilon}
\newcommand{\esp}[1]{\mathrm{e}^{#1}}    
\newcommand{\F}{f}              
\newcommand{\Fb}{\mathcal{F}}   
\newcommand{\Fh}{\CMcal{F}}     
\newcommand{\Fm}{\tilde{f}}     
\newcommand{\ff}{\mathcal{N}}              
\newcommand{\G}{\mathcal{G}}              
\newcommand{\gb}{\bar{\gamma}}  
\newcommand{\hyp}[4]{\hspace{0.5em}F\hspace{-1.15em}_2\hspace{0.65em}_1
  \left(\left.\begin{matrix}#1 \,, \; #2 \\ #3 \end{matrix}\right| #4 \right)}
\newcommand{\hypF}[2]{\hspace{0.5em}F\hspace{-1.15em}_{#1}\hspace{0.65em}_{#2}}
\newcommand{\K}{\mathcal{K}_{\e}}   
\newcommand{\Kc}{K^{\mathrm{coll}}} 
\newcommand{\kt}{\boldsymbol{k}}          
\newcommand{\MSbar}{\mathrm{MS}}          
\newcommand{\N}{\mathbb{N}}                
\newcommand{\om}{\omega}
\newcommand{\ord}[1]{\CMcal{O}\left(#1\right)}
\newcommand{\R}{\mathbb{R}}              
\newcommand{\regA}{{\it R1}}              
\newcommand{\regB}{{\it R2}}              
\newcommand{\regC}{{\it R3}}              
\newcommand{\rr}{\mathfrak{R}}           
\DeclareMathOperator{\sign}{sign}        
\newcommand{\ti}{T}                      
\newcommand{\ui}{\mathrm{i}}             
\newcommand{\z}{\zeta}

\begin{document}


\titlepage

\begin{center}
  {\Large \bf A solvable model for small-$\boldsymbol{x}$ physics\\[1ex]
    in $\boldsymbol{D>4}$ dimensions}\\
\bigskip\bigskip
  D.~Colferai\\
\bigskip\bigskip
  {\small\it  Dipartimento di Fisica, Universit\`a di Firenze,
   50019 Sesto Fiorentino (FI), Italy; \\
   INFN Sezione di Firenze,  50019 Sesto Fiorentino (FI), Italy}.\\
   E-mail: {\tt colferai@fi.infn.it}
\end{center}

\bigskip\bigskip


\begin{abstract}
  I present a simplified model for the gluon Green's function governing
  high-energy QCD dynamics, in arbitrary space-time dimensions. The BFKL
  integral equation (either with or without running coupling) reduces to a
  second order differential equation that can be solved in terms of Bessel and
  hypergeometric functions.  Explicit expressions for the gluon density and its
  anomalous dimension are derived in $\MSbar$ and $Q_0$ factorization
  schemes. This analysis illustrates the qualitative features of the QCD gluon
  density in both factorization schemes. In addition, it clarifies the
  mathematical properties and validates the results of the
  ``$\gamma$-representation'' method~\cite{dimensional} proposed by M.Ciafaloni
  and myself for extracting resummed next-to-leading-$\log x$ anomalous
  dimensions of phenomenological relevance in the two schemes.
\end{abstract}

\bigskip\bigskip


\bigskip\bigskip

\begin{minipage}{0.9\textwidth}
\begin{flushright}
  DFF 439/11/07\\
\end{flushright}
\end{minipage}

\newpage

\section{Introduction\label{s:intro}}

Small-$x$ resummations in QCD have been extensively investigated in the past
years in order to improve the fixed order perturbative description of
high-energy hard processes in the small-$x$ regime, where higher order
perturbative corrections grow rapidly due to logarithmically enhanced
contributions $\sim(\as\log1/x)^n$.
Knowledge of the precise relationship between the fixed order approach --- based
on the collinear factorization formula and the DGLAP equation\cite{DGLAP} ---
and the small-$x$ resummed ones --- based on the high-energy factorization
formula~\cite{Kfact} and the BFKL equation\cite{BFKL} --- is of course needed
for a unified picture of small-$x$ physics, e.g., to provide
quantitatively accurate predictions in the small-$x$ region, which will be
explored by next-generation colliders.

A major aspect of this relationship is the issue of the factorization scheme
employed to define parton densities and coefficient functions. Fixed order
perturbative calculations mostly use the (modified) minimal subtraction
($\MSbar$) scheme in the context of dimensional regularization.  On the other
hand, small-$x$ resummed approaches --- being based on
$\kt$-factorization~\cite{Kfact} which involves off-shell intermediate particles
with non-vanishing transverse momentum $\kt$ --- are naturally defined in the
so-called $Q_0$-scheme~\cite{Q0}, where infra-red (IR) singularities are
regularized by an off-shell probe whose non-vanishing virtuality $Q_0^2$ plays
the role of an IR cutoff.

The basic relations for the $\MSbar \leftrightarrow Q_0$ scheme change of
anomalous dimensions and coefficient functions were obtained some time
ago~\cite{CaCiHa93,CaHa94} at relative leading-$\log x$ (LL$x$) order, then
improved to include next-to-leading-$\log x$ (NL$x$) running coupling
corrections~\cite{CaCi97} and recently extended by M.C.~and myself at full NL$x$
level~\cite{dimensional}.  The main tool of our analysis~\cite{dimensional} was
the generalization to $4+2\e$ dimensions of the $\gamma$-representation of the
gluon density --- a Mellin representation of the BFKL solution in which $\gamma$
is conjugate to $t\equiv\log(\kt^2/\mu^2)$.  While for $\e=0$ the
running-coupling BFKL equation is a differential equation in $\gamma$, for
$\e\neq 0$ it becomes a finite-difference equation, whose solution, however, is
not unambiguously determined and has been computed by using sometimes rather
formal manipulations.  Despite the sensible physical meaning of the procedure
and of its results, from a mathematical point of view some steps of our method
are not fully proven. It is therefore desirable to have at least an explicit
example that could confirm our method of solution of the finite-difference
equation, especially in view of its application to compute the anomalous
dimensions at full NL$x$ accuracy.

The purpose of the present work is to devise a non-trivial, physically motivated
and solvable model which: {\it 1)} by providing explicit solutions, illustrates
the main qualitative features of the real QCD case; {\it 2)} can clarify the
less understood aspects of the procedure developed in~\cite{dimensional} and
verify the correctness of its results. The model I am going to present is a
generalization to arbitrary $D=4+2\e$ space-time dimensions of the collinear
model~\cite{collModel} used in the past to study the interplay between
perturbative and non-perturbative QCD dynamics at high energies. Starting from
the formulation of the LL$x$ BFKL equation in $D$ dimensions, only the
collinearly enhanced ($\kt \gg \kt'$ and $\kt \ll \kt'$) contributions of the
integral kernel $K(\kt,\kt')$ are kept. Despite its poor phenomenological
accuracy, this model contains most of the qualitative features of the real
theory: it is symmetric in the gluon exchange $\kt \leftrightarrow \kt'$, it
generates collinear singularities in the $\e \to 0$ limit, it correctly
describes the leading-twist LL$x$ behaviour of the gluon density, it includes
the running of the coupling. Most importantly, in contrast to the BFKL equation,
the collinear model can be solved, as a 1-dimensional Schr\"odinger-like
problem.

Sec.~\ref{s:fcm} is devoted to the definition of the model in generic number of
dimensions. A preliminary study on the qualitative features of the solution of
the master integral equation is presented. To this purpose, I briefly review the
two types of running-coupling behaviour that are present in $4+2\e$ dimensions.

The resolution of the model in the fixed-coupling case is presented in
sec.~\ref{s:b0}.  The ensuing integral equation is then recast into a second
order differential equation of Bessel type, whose solution provides the
unintegrated gluon density. The unintegrated gluon density is first compared
with the known perturbative solution~\cite{CaHa94} and then used to compute the
integrated gluon density and anomalous dimension in both the $\MSbar$-scheme and
$Q_0$-scheme. The last part of this section concerns the analysis of the Mellin
representation of the gluon density and its comparison with the corresponding
series and integral representations derived in ref.~\cite{dimensional}.

Sec.~\ref{s:bpos} includes the one-loop running coupling. In this case the
differential equation is solved in terms of hypergeometric functions. The
analyticity properties of the solution will reveal essential in extending the
unintegrated gluon density from the IR-free regime --- where the coupling is
bounded --- to the ultra-violet (UV)-free regime --- where the Landau pole
renders the integral equation meaningless.  The explicit results of the
$b$-dependent resummed $\MSbar$ and $Q_0$ anomalous dimensions --- which are
shown to agree with the known lowest order running coupling corrections ---
provide a strong check for the connection between $\e$-dependence of the kernel
and $b$-dependence of the $\MSbar$ anomalous dimension argued in
ref.~\cite{dimensional}.

A final discussion is reported in sec.~\ref{s:c}.

\subsection{Notations\label{ss:notations}}

I distinguish two symbols of asymptotic behaviour:
$f(x) \sim g(x)$ means $\lim_{x\to x_0} \frac{f(x)}{g(x)} = k$ for some finite
and non-zero $k$, while $f(x) \approx g(x)$ refers to the special case $k = 1$.\\
The hypergeometric function is denoted by
$\hypF{2}{1}(a,b;c;z) \equiv \hyp{a}{b}{c}{z}$.\\
A citation like [1](2.3) means eq.~(2.3) of ref.~[1].\\
There are some change of notations between
ref.~\cite{dimensional} (left side) and this paper (right side):
\begin{align*}
  \CMcal{F}(\kt) &\quad\longrightarrow\quad \om \G(\kt,\kt_0) \\
  \widetilde{\CMcal{F}}_{\e}(\kt) \, \esp{\e\psi(1)} / \Gamma(1+\e)
 &\quad\longrightarrow\quad \F_{\e}(t) \\
  -T &\quad\longrightarrow\quad t_0 \;.
\end{align*}

\section{Formulation of the collinear model\label{s:fcm}}

In this section, I define a simplified model for the gluon density in
high-energy QCD with both running and frozen coupling constant. After recalling
the features of the two running coupling regimes, I briefly discuss the expected
qualitative behaviour of the solutions of the model.

\subsection{Motivation of the model\label{ss:mm}}

In high-energy QCD, parton densities and anomalous dimensions are often
computed in two different factorization schemes, which differ essentially by the
regularization of the infra-red (IR) singularities.
\begin{itemize}
\item In the so-called $Q_0$ scheme~\cite{Q0}, the IR regularization occurs by
  considering off-shell initial partons with non-vanishing virtuality
  $k^2 = -Q_0^2 < 0$, which plays the role of a momentum cut-off;
\item The minimal subtraction ($\MSbar$) scheme instead, is based on dimensional
  regularization with on-shell initial partons living in $D=4+2\e$ space-time
  dimensions, where IR singularities shows up as poles $\sim 1/\e^n$ and are
  subtracted from the physical quantities according to the $\MSbar$ prescription.
\end{itemize}
The relation between these two schemes can be investigated by including in the
defining equations for partons both off-shell initial conditions {\em and}
arbitrary space-time dimensions.

As for the physical case of 4 space-time dimensions, also in generic $D=4+2\e$
dimensions the high energy (i.e., small-$x$) behaviour of cross sections in QCD
is governed by the gluon Green's function (GGF) $\G_{\om,\e}(\kt,\kt_0)$. Here
$\om$ is the Mellin variable conjugated to $x$, while $\kt$ and $\kt_0$ are the
transverse momenta of the (reggeized) gluons emerging from the impact-factors of
the external particles~\cite{Kfact}. The GGF obeys the integral equation (in the
following the dependence on the $\om$ variable will always be understood)
\begin{equation}\label{intEqKt}
 \om \G_{\e}(\kt,\kt_0) = \delta^{2+2\e}(\kt-\kt_0) +
 \int\frac{\dif^{2+2\e}\kt'}{(2\pi)^{2+2\e}} \;
 \K(\kt,\kt') \G_{\e}(\kt',\kt_0)
\end{equation}
where the kernel $\K$ has been determined exactly in the leading-$\log(x)$
(LL$x$) approximation~\cite{CaHa94} and can be conveniently improved to
include subleading corrections (in particular the running of the coupling).
Detailed studies of the ensuing solutions and physical consequences has been
presented in refs.~\cite{CaHa94} in the LL$x$ approximation, and in
refs.~\cite{dimensional,CCSS06} at subleading level.

It should be noted that the NL$x$ approximation limits not only the knowledge of
the kernel $\K$, but also the method of solution of eq.~(\ref{intEqKt}).
However, it would be desirable to have an exact solution of eq.~(\ref{intEqKt}),
even with an approximate kernel, in order to understand the overall
``non-perturbative'' feature of the QCD gluon Green's function. To this purpose,
I consider a simplified model for $\K$ whose main virtue is to provide a GGF
which can be expressed in terms of known analytic functions. This toy-kernel
resembles the field-theoretical one in the collinear regions $\kt^2 \ll
\kt'{}^2$ and $\kt^2 \gg \kt'{}^2$, and has already been considered in the
past~\cite{collModel} in order to study the structure of high-energy QCD
dynamics in 4 dimensions. In the following I generalize the collinear model to
the dimensional regularized theory, with running coupling as well as with fixed
coupling constant.

\subsection{Definition of the model\label{ss:dm}}

The collinear model is defined by the collinear limit $\kt_<^2 \ll \kt_>^2$ of
the LL$x$ BFKL high-energy evolution kernel
\begin{equation}\label{collLimit}
 \K^{\mathrm{BFKL}}(\kt,\kt') = \frac{g^2 N_c}{\pi(\kt-\kt')^2}
 + \text{virtual terms} \quad\longrightarrow\quad
 \frac{g^2 N_c}{\pi\kt_>^2} \equiv \K^{\mathrm{coll}}(\kt,\kt') \;,
\end{equation}
$\kt_<$ ($\kt_>$) being the smallest (biggest) transverse momentum, and $g$ the
dimensionful gauge coupling.  By introducing the dimensionless
coupling constant $\asb$, the small-$x$ expansion parameter $a$ and the
logarithmic variable $t$
\begin{equation}\label{d:asb}
 \asb \equiv \frac{(g\mu^\e)^2}{(4\pi)^{1+\e}\Gamma(1+\e)}\frac{N_c}{\pi}
 \;, \qquad
 a \equiv \frac{\asb}{\om}
 \;, \qquad
 t \equiv \log\frac{\kt^2}{\mu^2} \;, 
\end{equation}
we can express both GGF and kernel in terms of the dimensionless quantities
$\F_{\e}$ (the unintegrated gluon density) and $\Kc$ defined by
\begin{align}
 \om\G_{\e}(\kt,\kt_0) &\equiv \delta^{2+2\e}(\kt-\kt_0)
  + \frac{\Gamma(1+\e)}{(\pi\kt^2)^{1+\e}} \F_{\e}(t,t_0)
 \label{d:F} \\
 \kt^2 \, \K^{\mathrm{coll}}(\kt,\kt') &\equiv \frac{g^2 N_c}{\pi} \Kc(t-t') \;.
 \label{d:K}
\end{align}
so that one can rewrite eq.~(\ref{intEqKt}) in the form%
\footnote{In ref.~\cite{dimensional} we adopted a step function $\Theta(t+T)$,
  $T\equiv -t_0$, instead of $\Kc(t-t_0)$ as inhomogeneous term.}
\begin{align}
 \F_{\e}(t,t_0) &= a \esp{\e t} \left[
  \Kc(t-t_0) + \int_{-\infty}^{+\infty} \Kc(t-t') \F(t',t_0) \; \dif t' \right]
 \label{intEqGen} \\
 &= a \esp{\e t} \left[ \Theta(t_0-t)\esp{t-t_0} + \Theta(t-t_0)
 + \int_{-\infty}^{t} \F(t',t_0) \,\dif t' + \int_t^{+\infty} \esp{t-t'}
 \F(t',t_0) \,\dif t' \right] \;,
 \label{intEqColl}
\end{align}
where in the second line I have substituted the expression of the collinear
kernel
\begin{equation}\label{Kcoll}
 \Kc(\tau) = \Theta(-\tau)\esp{\tau} + \Theta(\tau) \;, \quad
 \tau \equiv t-t'
\end{equation}
stemming from eqs.~(\ref{collLimit}) and (\ref{d:K}).

\begin{figure}[hb]
  \centering
  \includegraphics[width=0.5\textwidth]{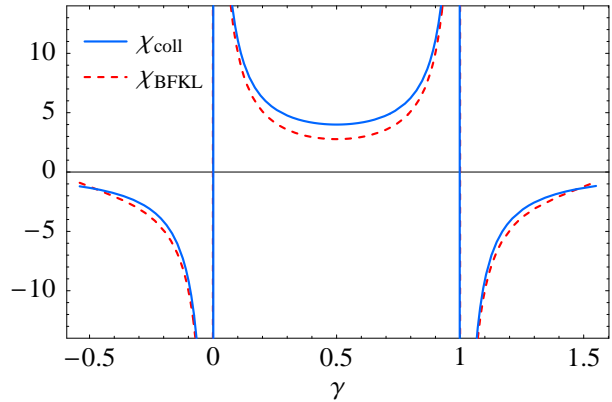}
  \caption{\sl Comparison of the collinear model eigenvalue function
    (solid-blue) with the BFKL one (dashed-red).}
  \label{f:chi}
\end{figure}

A second way to relate this model to QCD is to compare the eigenvalue function
\begin{equation}\label{chi}
 \chi_{\mathrm{coll}}(\gamma) \equiv \chi(\gamma)
 \equiv \int_{-\infty}^{+\infty} \esp{-\gamma\tau}
 \Kc(\tau) \, \dif\tau = \frac1{\gamma} + \frac1{1-\gamma} \;,
\end{equation}
with the BFKL one $\chi_{\mathrm{BFKL}} = 2\psi(1) -\psi(\gamma) -
\psi(1-\gamma)$, as in fig.~\ref{f:chi}. Clearly the two eigenvalue functions
display the same qualitative behaviour in the region around and between the
leading-twist poles at $\gamma = 0, 1$.

The collinear model can be easily generalized to include the running of the
coupling. The small-$x$ parameter $a$ acquires a $t$-dependence according to the
evolution equation
\begin{equation}\label{rce}
 \frac{\dif a(t)}{\dif t} = \e [a(t) - B a^2(t)] \;, \qquad
 B \equiv \frac{b\om}{\e}
\end{equation}
where $b$ is the one-loop beta function coefficient
($b = 11/12 - N_f/6 N_c$ in QCD). The solution of eq.~(\ref{rce}) is given by
\begin{equation}\label{rc}
 a_t \equiv a(t) = \frac{a\esp{\e t}}{1 + a B (\esp{\e t} - 1)} 
 \qquad \iff \qquad \left(\frac1{a(t)}-B\right)^{-1} = A\esp{\e t} \;, \quad
 A \equiv \frac{a}{1-aB} \;.
\end{equation}
Note that in dimensional regularization ($\e \neq 0$) the coupling $a(t)$ has a
non-trivial $t$-dependence also in the so-called frozen coupling case
corresponding to $b = 0$.  Substituting $a(t)$ in place of $a\esp{\e t}$ in
eq.~(\ref{intEqGen}), we obtain, after rearranging some terms, the
generalization of the collinear model with running coupling:
\begin{align}
 &\F_{\e,b}(t,t_0) = A\esp{\e t} \left[ \Kc(t-t_0) +
    \int_{-\infty}^{+\infty} \Kc(t-t',B) \F_{\e,b}(t',t_0) \; \dif t' \right]
 \label{intEqRc} \\
 & \Kc(\tau,B) \equiv \Kc(\tau) - B \delta(\tau) \;.
 \label{d:KcB}
\end{align}

\subsection{Running coupling regimes\label{ss:rcr}}

\begin{figure}[!t]
  \centering
  \includegraphics[width=0.5\textwidth]{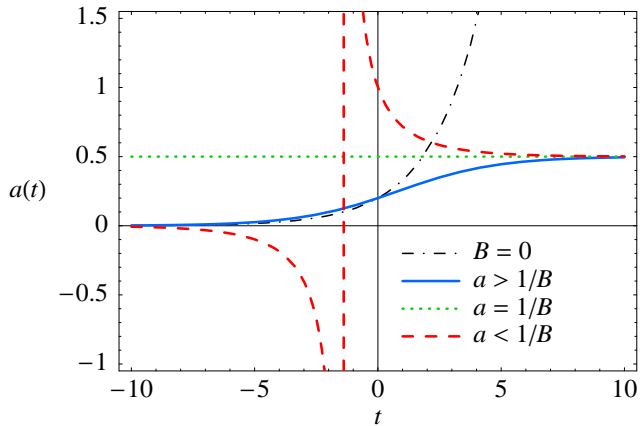}
  \caption{\sl Behaviour of the running coupling $a(t)$ in the regular regime $aB <
    1$ (solid-red) and in the Landau regime $aB > 1$ (dashed-blue). The straight
  line (dotted-green) corresponds to the boundary value $aB = 1$. The case $B=0$
  is represented by the dash-dotted black curve.}
  \label{f:rc}
\end{figure}

It is important at this point to realize that the running coupling behaves in
two qualitatively different ways, according to whether the parameter $aB$ is
greater or less than 1.
\begin{itemize}
\item When $aB < 1$, i.e., $\asb < \e/b$, the running coupling $a(t)$ is
  bounded, positive and increases monotonically from the IR-stable fixed point
  $a(-\infty) = 0$ to the UV-stable fixed point $a(+\infty) = 1/B$, as shown in
  fig.~\ref{f:rc}.
\item When $aB > 1$, i.e., $\asb > \e/b$, the running coupling starts from the
  positive UV-stable fixed point $a(+\infty) = 1/B$, then increases and diverges
  at the Landau point
  \begin{equation}\label{landau}
    t_\Lambda \equiv -\frac1{\e}\log(-AB)
    = \frac1{\e}\log\left(1-\frac{\e}{ab}\right) \;,
  \end{equation}
  becomes negative for $t < t_\Lambda$ and finally vanishes at $t = -\infty$.
  This is the situation realizing the physical limit $\e \to 0$ at fixed $b$.
\end{itemize}
In the former case, the extra-dimension parameter $\e$ not only regularizes the
IR singularities, but avoids also the occurrence of the Landau pole, thus
allowing a formulation of the integral equation free of singularities. In
practice, the strategy of dimensional regularization consists in computing the
physical quantities in the ``regular'' regime $\asb < \e/b$; the universal
$\e$-singular factors are then removed into non-perturbative quantities, and
finally by analytic continuation the physical case at $\e = 0$ is recovered.

\subsection{Qualitative behaviour of the solutions\label{ss:qbs}}

Before embarking upon the resolution of the collinear model
equations~(\ref{intEqColl},\ref{intEqRc}), it is instructive to estimate the
qualitative behaviour of the solutions by using well-known methods~\cite{Q0} in
the context of high-energy QCD. Particularly important is the factorization
property which allows one to split the unintegrated gluon density $\F(t,t_0)$
into a perturbative and a non-perturbative part, provided the ``hard scale''
$t \gg t_0 \gtrsim t_\Lambda$ is sufficiently large:
\begin{equation}\label{QCDfact}
  \F(t,t_0) = \F_{\mathrm{pt}}(t) \F_{\mathrm{np}}(t_0) \times
  [1+\ord{\esp{-t}}] \;,
\end{equation}
up to terms exponentially suppressed in $t$ (higher-twists). In turn, the
perturbative factor
\begin{equation}\label{pertFact}
 \F_{\mathrm{pt}}(t) \sim \exp\left\{\int^t \gb\big(a(t')\big) \;\dif t'\right\}
\end{equation}
is given in terms of the gluon anomalous dimension
$\gb\big(a(t)\big)$ determined by the small-$x$ equation
\begin{equation}\label{sxEq}
  1 = a(t) \chi(\gb) \;,
\end{equation}
where $\chi$ is the eigenvalue function of the integral kernel in
eq.~(\ref{intEqKt}).

In this collinear model, the eigenvalue function in eq.~(\ref{chi}) provides two
solution to eq.~(\ref{sxEq})
\begin{equation}\label{gb}
  \gb_{\pm}(a) = \frac{1\pm\sqrt{1-4a}}{2} \;,
\end{equation}
the perturbative branch being the one with minus sign:
$\gb_{-}(a) = a + \ord{a^2}$.  At large $t$, the running coupling saturates at
the UV fixed point $a(+\infty) = 1/B = \e/b\om$, so that the large-$t$ behaviour
of the unintegrated gluon density is given by
\begin{equation}\label{fUV}
 \F(t \gg t_0)
 \sim \sum_{j=\pm} c_j(t_0) \esp{t \; \gb_j\left(a(+\infty)\right)}
 = c_+(t_0) \esp{\frac{t}{2}\left(1+\sqrt{1-\frac{4}{B}}\right)}
 + c_-(t_0) \esp{\frac{t}{2}\left(1-\sqrt{1-\frac{4}{B}}\right)} \;.
\end{equation}
According to the value of $B$ we expect two kinds of asymptotic behaviour:
\begin{itemize}
\item For $B>4$ the square root is real and positive, the UV regular solution
  corresponds to the perturbative branch $\gb_-$ of the anomalous dimension and
  we must reject the (UV irregular) solution which diverges more rapidly:
  $c_+=0$.
\item For $B<4$ the two exponents are complex conjugate, and the gluon density
  becomes oscillatory at large $t$. It is not possible to distinguish an UV
  regular solution, and one has to determine the coefficients $c_{\pm}$ by
  analytic continuation in $B$ from $B > 4$.  The fixed coupling ($B=0$)
  solution belongs to this class.
\end{itemize}
The above results will be also obtained in a more rigorous way in
sec.~\ref{ss:abs}, when treating the running-coupling equation.

\section{Collinear model with frozen coupling ($\boldsymbol{b = 0}$)
  \label{s:b0}}

Since the properties of the solution of the collinear model and its connection
with the solution method of ref.~\cite{dimensional} are more easily illustrated
in the fixed coupling case, I start considering the integral
equation~(\ref{intEqColl}) with $b = 0$.

\subsection{Solution in momentum space\label{ss:sms}}

The presence of the exponential factor in front of the r.h.s.\ of
eqs.~(\ref{intEqGen},\ref{intEqColl}) spoils scale invariance, therefore the
determination of both eigenfunctions and eigenvalues of the integral operator by
means of standard techniques is not possible.  It turns out, however, that one
can exactly solve eq.~(\ref{intEqColl}). In fact, by differentiating it twice
with respect to $t$, we obtain the second order differential equation (the
$\e$-dependence of $\F$ is understood in this section)
\begin{equation}\label{diffEq}
 \F''-(1+2\e)\F'+[\e(1+\e)+a\esp{\e t}]\F = -a\,\esp{\e t_0}\,\delta(t-t_0)\;,
\end{equation}
which can be recast in a more familiar form if we introduce the variables
\begin{equation}\label{besselVar}
 \eta \equiv \frac1{\e} \;, \quad z \equiv 2\eta\sqrt{a \esp{\e t}}
 \;, \quad \F(t,t_0) \equiv z^{\eta+2} \Fb(z,z_0) \;,
\end{equation}
thus obtaining
\begin{equation}\label{besselDiffEq}
 z^2 \Fb'' + z \Fb' + [z^2 - \eta^2]\Fb =
 -\frac{\delta(z-z_0)}{2\eta z_0^{\eta-1}} \;.
\end{equation}
In the l.h.s.\ of eq.~(\ref{besselDiffEq}) one recognizes the differential
operator defining the Bessel functions $J_{\pm\eta}(z)$ and $Y_\eta(z)$ as
solutions of the corresponding homogeneous equation.

The general solution of eq.~(\ref{besselDiffEq}) has the form
\begin{equation}\label{besselGenSol}
 \Fb(z,z_0) =  c_I(z_0) \Fb_I(z) \Theta(z_0-z) + c_U(z_0) \Fb_U(z) \Theta(z-z_0)
\end{equation}
where $\Fb_I$ and $\Fb_U$ denote respectively the IR-regular and the UV-regular
solutions of the homogeneous equation, while $c_I$ and $c_U$ are $z_0$-dependent
coefficients to be determined by the two conditions of continuity of $\Fb$ and
discontinuity of $\partial_z\Fb$ at $z = z_0$:
\begin{subequations}\label{matching}
\begin{align}
  &\lim_{z\to z_0^+} \Fb(z,z_0) - \lim_{z\to z_0^-} \Fb(z,z_0)
  &=&& c_U(z_0) \Fb_U(z_0) - c_I(z_0) \Fb_I(z_0) 
  &&=& \quad 0 \\
  &\lim_{z\to z_0^+} \partial_z\Fb(z,z_0) - \lim_{z\to z_0^-}
  \partial_z\Fb(z,z_0)
  &=&& c_U(z_0) \Fb'_U(z_0) - c_I(z_0) \Fb'_I(z_0) 
  &&=& \quad -\frac1{2\eta z_0^{\eta+1}} \equiv -N(z_0) \;.
\end{align}
\end{subequations}
By solving the above linear system one obtains
\begin{equation}\label{besselSymSol}
  \Fb(z,z_0) = \frac{N(z_0)}{W(z_0)} \left[ \Fb_I(z) \Fb_U(z_0) \Theta(z_0-z) +
    \Fb_U(z) \Fb_I(z_0) \Theta(z-z_0) \right] \;,
\end{equation}
where $W = \Fb_U \Fb'_I - \Fb'_U \Fb_I$ is the Wronskian of the two solutions of
the homogeneous equation.

It remains to determine $\Fb_I$ and $\Fb_U$, each being a linear combinations
of, say, $J_\eta$ and $Y_\eta$:
\begin{equation}\label{besselLinComb}
  \Fb_s(z) = c_s^{(1)} J_\eta(z) + c_s^{(2)} Y_\eta(z) \;, \qquad ( s = I, U)
\end{equation}
(the absolute normalization is irrelevant).  From the asymptotic relations
\begin{subequations}\label{besselAsym}
  \begin{align}
    J_\eta(z) &\sim z^\eta \;\;\qquad (z\to 0)
    & J_\eta(z) &\sim z^{-1/2}\cos(z+\phi_1) \qquad (z \to +\infty) \\
    Y_\eta(z) &\sim z^{-\eta} \qquad (z\to 0)
    & Y_\eta(z) &\sim z^{-1/2}\cos(z+\phi_2) \qquad (z \to +\infty)
  \end{align}
\end{subequations}
it is clear that the IR-regular solution is $\Fb_I \propto J_\eta$, since it
vanishes more rapidly than any linear combination containing $Y_\eta(z)$ when
$z \to 0$ with $\eta > 0$.  On the other hand, the UV-regular solution cannot be
determined in this case of $b = 0$, because of the identical asymptotic
behaviour (up to normalization and phase) for $z \to +\infty$ of all solutions
in eq.~(\ref{besselLinComb}).  However, the UV-regular solution can be
unambiguously determined in the formulation with running coupling
(cf.~sec.~\ref{ss:fcl}), and in the $b \to 0$ limit it reduces to
$\Fb_U(z) = Y_\eta(z)$. In conclusion
\begin{align}
  \Fb_I(z) &= J_\eta(z)
 \label{FbI} \\
  \Fb_U(z) &= Y_\eta(z) = [\cos(\pi\eta) J_\eta(z) - J_{-\eta}(z)] / \sin(\pi\eta)
 \label{FbU} \\
  W(z) &= Y_\eta J'_\eta - J_\eta Y'_\eta = -2/\pi z \;,
 \label{Wb}
\end{align}
whence
\begin{equation}\label{besselSol}
 \Fb(z,z_0) = -\frac{\pi}{4\eta z_0^\eta} \left[ J_\eta(z) Y_\eta(z_0) \Theta(z_0-z) +
 Y_\eta(z) J_\eta(z_0) \Theta(z-z_0) \right] \;.
\end{equation}
It is possible to show that $\Fb(z,z_0)$ in the previous equation obeys also the
integral equation~(\ref{intEqColl}).

\subsection{On-shell limit and perturbative expansion\label{ss:oslpe}}

It is important at this point to check the explicit solution in
eq.~(\ref{besselSol}) with known results of the literature.  The perturbative
expression for the GGF $\G(\kt,\kt_0)$ in dimensional regularization was given
in~\cite{CaHa94}(3.3) for an on-shell ($\kt_0 = 0$) initial gluon. In terms
of the dimensionless density $\F_\e$ their result reads
\begin{equation}\label{pertSol}
 \F_\e(t) = a\esp{\e t}\left[1+\sum_{m=1}^\infty \left(a\esp{\e t}\right)^m
 \prod_{j=1}^m \chi(j\e,\e) \right]
\end{equation}
for a generic integral kernel with eigenvalue function $\chi(\gamma,\e)$.

The on-shell limit $\kt_0 \to 0 \iff t_0 \to -\infty$ of the unintegrated gluon
density $\F_\e(t,t_0)$ at fixed $t,\,\e,\,\asb$ is finite, and can be obtained
from eqs.~(\ref{besselVar},\ref{besselSol}) by
exploiting the asymptotic behaviour of Bessel functions
$J_{\eta}(z_0) \approx (z_0/2)^{\eta}/\Gamma(1+\eta)$ for $z_0\to0$, whence
\begin{equation}\label{osBesselSol}
 \F_{\e}(t) \equiv \lim_{t_0 \to -\infty} \F_{\e}(t,t_0)
 = z^{\eta+2}\lim_{z_0 \to 0^+}-\frac{\pi}{4\eta z_0^\eta} J_\eta(z_0) Y_\eta(z)
 = -\frac{\pi}{\eta\Gamma(\eta+1)} \left(\frac{z}{2}\right)^{\eta+2}Y_\eta(z)\;.
\end{equation}
In words, the on-shell unintegrated gluon density is equal to the UV regular
solution of the homogeneous {\em differential} equation with a proper
normalization.

In order to compare the solution~(\ref{osBesselSol}) with the perturbative
expression~(\ref{pertSol}), one has to expand the r.h.s.\ of
eq.~(\ref{osBesselSol}) in series of $z^2 \sim a$. By rewriting $Y_\eta$ as a
combination of $J_{\pm\eta}$ according to eq.~(\ref{FbU}), and then using the
ascending series~\cite{AS}(9.1.10)
\begin{equation}\label{besselSeries}
  J_\nu(z) = \left(\frac{z}{2}\right)^\nu \sum_{m=0}^\infty
 \frac{(-z^2/4)^m}{m! \Gamma(1+\nu+m)} \;,
\end{equation}
one obtains
\begin{align} \label{besselSolSeries}
 \F_{\e}(t) &= \frac{\pi}{\eta\Gamma(\eta+1)\sin(\pi\eta)}
 \left(\frac{z}{2}\right)^{\eta+2}
 \left[ J_{-\eta}(z) - \cos(\pi\eta) J_\eta(z) \right] \\ \nonumber
 &= a\esp{\e t} \sum_{m=0}^\infty
 \frac{\left(-\eta^2 a\esp{\e t}\right)^m}{m!}\frac{\Gamma(1-\eta)}
 {\Gamma(1-\eta+m)}
 - \cos(\pi\eta)\Gamma(1-\eta) \eta^{2\eta} \left(a\esp{\e t}\right)^{\eta+1}
 \sum_{m=0}^\infty \frac{\left(-\eta^2 a\esp{\e t}\right)^m}{m!
   \Gamma(1+\eta+m)} \;.
\end{align}
The first term in the r.h.s.\ of eq.~(\ref{besselSolSeries}) exactly reproduces
the perturbative result~(\ref{pertSol}), since for $m \geq 1$
\begin{equation}\label{pertCoef}
 \frac{-\eta^{2m}}{m!}\frac{\Gamma(1-\eta)}{\Gamma(1-\eta+m)}
 = \prod_{j=1}^m \frac{-1/\e^2}{j (-\frac1{\e}+j)}
 = \prod_{j=1}^m \frac1{j\e(1-j\e)} = \prod_{j=1}^m \chi(j\e) \;.
\end{equation}
The second term of eq.~(\ref{besselSolSeries}) provides contributions of order
$(a\esp{\e t})^{\eta+1+m} = a^{1+m+1/\e} \esp{[1+(1+m)\e]t}$, each being outside
the domain of the kernel and therefore out of the reach of the iterative
procedure. Furthermore, this term is strongly suppressed $\sim a^{\frac1{\e}}$
when $\e\to0$ with respect to the perturbative one.  Therefore, it is possible
to correctly compute the perturbative coefficients to any order $m$ provided
$\e$ is sufficiently small ($\e < 1/m$). In the limit $\e \to 0$ the
perturbative solution agrees with the exact one to all orders.%
\footnote{These conclusions are valid in the off-shell case ($t_0\in\R$) too,
  but for sake of simplicity they have been presented only in the on-shell
  case.}

As final remark, the series in eqs.~(\ref{pertSol},\ref{besselSolSeries})
converge for all $t\in\R$, as one can check from the $\gamma\to+\infty$
asymptotic behaviour of $\chi(\gamma) \sim 1/\gamma$.

\subsection{Integrated gluon densities\label{ss:igd}}

The major issue this paper is devoted to, concerns the
$\MSbar\leftrightarrow Q_0$ scheme-change, namely the relation between gluon
densities and anomalous dimensions in the two factorization schemes.  In the
collinear model, the off-shell {\em integrated} gluon density defined by
\begin{equation}\label{d:gluon}
 g_{\e}(t,t_0) \equiv \int\dif^{2+2\e}\kt' \; \om \G(\kt',\kt_0)
 \Theta(\kt^2-\kt'{}^2) = 1 + \int_{-\infty}^t \dif t'\; \F_{\e}(t',t_0)
\end{equation}
can be computed in closed form (app.~\ref{a:igdb0}), and for $t > t_0$ reads
\begin{equation}\label{gluon}
 g_{\e}(t,t_0) = -\pi \, \frac{z}{2} \left(\frac{z}{z_0}\right)^\eta
 J_{\eta}(z_0) Y_{\eta+1}(z) \qquad (t > t_0) \;.
\end{equation}
Note the remarkable fact that $g$, like $\F$, is factorized in the $t$- and
$t_0$-dependence.

The {\em $Q_0$-scheme} gluon is given by the $\e \to 0$ limit of the above
expression, yielding (app.~\ref{a:igdb0e0})
\begin{equation}\label{gQ0}
 g^{(Q_0)}(t,t_0) \equiv \lim_{\e\to0}g_{\e}(t,t_0)
 = \frac{a}{\sqrt{1-4a}\,\gb(a)} \exp\left[\gb(a)(t-t_0)\right]
\end{equation}
whence one immediately derives the $Q_0$-scheme anomalous dimension (a dot
means $t$-derivative)
\begin{equation}\label{gammaQ0}
 \gamma^{(Q_0)}(a) \equiv \lim_{t_0 \to -\infty}
 \frac{\dot{g}^{(Q_0)}(t,t_0)}{g^{(Q_0)}(t,t_0)} = \gb(a)
 \qquad (\dot{g}\equiv\partial_{t}g) \;.
\end{equation}
It is interesting to note that the on-shell limit of the integrated gluon
density, i.e., the gluon density in dimensional regularization
\begin{equation}\label{osGluon}
 g_{\e}(t) \equiv \lim_{t_0 \to -\infty} g_{\e}(t,t_0)
 = -\frac{\pi}{\Gamma(\eta+1)} \left(\frac{z}{2}\right)^{\eta+1} Y_{\eta+1}(z)
\end{equation}
provides the same {\em effective anomalous dimension} (app.~\ref{a:igdb0e0})
\begin{equation}\label{gammaEps}
 \gamma_{\mathrm{eff}}(t) \equiv \lim_{\e\to0} \frac{\dot{g}_{\e}(t)}{g_{\e}(t)}
 = \lim_{\e\to0}\frac{\F_{\e}(t)}{g_{\e}(t)}
 = \lim_{\e\to0} \frac{z}{2\eta}\frac{Y_\eta(z)}{Y_{\eta+1}(z)}
 = \gb(a) \;,
\end{equation}
which means that the two limiting operations $t_0 \to -\infty$ and $\e \to 0$
commute. Actually, in this model this is a trivial consequence of the factorized
structure of the gluon density~(\ref{gluon}) in its $t$ and $t_0$ dependence,
which causes the ratios $\dot{g}/g$ in eqs.~(\ref{gammaQ0}) and (\ref{gammaEps})
to be $t_0$-independent.

The relation with the {\em MS-scheme} anomalous dimension is obtained as
follows. From the $\e \to 0$ asymptotic behaviour of the on-shell gluon density
(app.~\ref{a:igdb0e0})
\begin{equation}\label{gAsy}
 g_{\e}(t) = [1+\ord{\e}]\; \frac{a}{\gb(a) [1-4a]^{1/4}}
 \exp\left\{\frac1{\e}\int_0^{a\esp{\e t}} \frac{\dif \aaa}{\aaa}\;
 \gb(\aaa) \right\} \equiv R_\e(a) g_\e^{(\MSbar)}(t)
\end{equation}
one identifies the exponential in eq.~(\ref{gAsy}) as the
$\MSbar$ gluon density $g^{(\MSbar)}(t)$,%
\footnote{Due to the particular definition of $\asb$ in eq.~(\ref{d:asb}) which
  includes $\e$-dependent factors, eq.~(\ref{gAsy}) defines a ``modified''
  minimal subtraction scheme, related to the customary MS and
  $\overline{\mathrm{MS}}$ schemes by a finite scheme change. These details are
  unimportant for the purpose of this paper.}
since it sums all and only $\e$-singular terms up to the scale
$\kt^2 = \mu^2\esp{\e t}$.  The $\MSbar$ anomalous dimension is then computed
from the logarithmic derivative
\begin{equation}\label{gammaMS}
 \gamma^{(\MSbar)}(a) \equiv \lim_{\e \to 0}
 \frac{\dot{g}^{(\MSbar)}_{\e}(t)}{g^{(\MSbar)}_{\e}(t)} = \gb(a)
\end{equation}
and coincides, in this case of $b=0$, with the $Q_0$-scheme anomalous dimension,
in agreement with refs.~\cite{CaHa94} and~\cite{dimensional}.

The coefficient function $R$, on the other hand, is finite in the $\e\to 0$
limit, and is given by the product~\cite{dimensional} $R = \ff \rr$, where
\begin{equation}\label{ffN}
 \ff(a) = \frac1{\gb(a) \sqrt{-\chi'\big(\gb(a)\big)}}
 = \frac{a}{\gb(a)[1-4a]^{1/4}}
\end{equation}
is the fluctuation factor of the saddle-point estimate introduced
in~\cite{dimensional} (cf.~also sec.~\ref{ss:sgs}), while
\begin{equation}\label{rr}
  \rr(a) = \exp\left\{\int_0^{\gb(a)}
 \frac{\chi_1(\gamma)}{\chi_0(\gamma)} \; \dif\gamma \right\} \;, \qquad
 \chi(\gamma,\e) = \chi_0(\gamma) + \e \chi_1(\gamma) + \ord{\e^2}
\end{equation}
originates from the $\e$-dependence of the eigenvalue function. Since in this
model $\chi$ is independent of $\e$, $\chi_1 = 0$, $\rr = 1$ and therefore
$R=\ff$, in agreement with eq.~(\ref{gAsy}).

\subsection{Solution in $\boldsymbol{\gamma}$ space\label{ss:sgs}}

Having the solution of the integral equation at our disposal, we are ready to
check the validity of the procedure suggested in ref.~\cite{dimensional}, at
least in this simplified model. I start reviewing the main steps of that
procedure.

{\bf 1)}
We introduce for the unintegrated gluon density $\F_{\e}(t,t_0)$ an integral
representation of Mellin-type:
\begin{equation}\label{intRep}
 \F_{\e}(t,t_0) = \int_{\contour} \frac{\dif \gamma}{2\pi\ui} \;
 \esp{\gamma t} \Fm_{\e}(\gamma,t_0) \;.
\end{equation}

{\bf 2)} In $\gamma$-space, the integral equation~(\ref{intEqColl}) is thus
recast into the finite difference equation
\begin{equation}\label{finDifEq}
 \Fm_{\e}(\gamma+\e,t_0) = a \chi(\gamma) \esp{-\gamma t_0}
 + a \chi(\gamma) \Fm_{\e}(\gamma,t_0) \;.
\end{equation}

{\bf 3)} The finite difference equation~(\ref{finDifEq}) is solved in terms of
a Laurent series in $\e$, so as to provide the following expression
(cf.~\cite{dimensional}, sec.~2 and eqs.~(C.1,C.2)) for the on-shell
unintegrated gluon density:
\begin{align}\label{marcelloSol}
 \F_\e(t) &= \Omega(a,\e) \int_{\contour} \frac{\dif \gamma}{2\pi\ui} \;
 \esp{\gamma t} \exp\left\{\frac1{\e} \int_0^\gamma L(\gamma') \; \dif\gamma'
 -\frac12 L(\gamma)
 + \sum_{n=2}^\infty \frac{B_n}{n!}\,\e^{n-1} L^{(n-1)}(\gamma) \right\} \;,
\end{align}
where $\Omega = \sqrt{a/2\pi\e}[1+\ord{\e}]$ is a normalization factor,
$L(\gamma) \equiv \log\big(a\chi(\gamma)\big)$,
$L^{(n)} \equiv \partial_\gamma^n L$, and
the coefficients $B_n$ denote Bernoulli numbers.

{\bf 4)} The solution is determined by assuming the existence of a saddle point
on the real axis, whose steepest descent direction lies on the real axis.\\

Let us now analyze each point in turn, in the context of the collinear model.

{\bf 1)} Concerning the existence of a Mellin representation for the solution
$\F$ of the integral equation~(\ref{intEqColl}), the asymptotics in
eq.~(\ref{besselAsym}) guarantee that the Mellin transform $\Fm$ is defined in
the strip $1/2+3\e/4 < \Re \gamma < 1+\e$ for all $\e > 0$. Explicitly, $\Fm$ is
given in terms of $\hypF{1}{2}(u;u+1,v;-z_0^2/4)$ sums, as follows:
\begin{subequations}\label{besselMellin}
\begin{align}
 \Fm_\e(\gamma,t_0) &\equiv \left( \int_{-\infty}^{t_0} + \int_{t_0}^{+\infty}
  \right) \dif t \; \esp{-\gamma t} \F_\e(t,t_0) \equiv \Fm_\e^{(-)}(\gamma,t_0)
 + \Fm_\e^{(+)}(\gamma,t_0)
 \label{d:Mellin} \\
  \Fm_\e^{(-)}(\gamma,t_0) &= \pi \left(\frac{z_0}{2}\right)^{\eta}
  Y_{\eta}(z_0) \,\esp{-\gamma t_0} \sum_{k=1}^\infty
 \frac{(-z_0^2/4)^k}{(k-1)! \, [k+\eta(1-\gamma)] \, \Gamma(k+\eta)}
 \label{besselMellinMinus} \\
 \Fm_\e^{(+)}(\gamma,t_0) &= \pi \left(\frac{z_0}{2}\right)^{-\eta}
 J_{\eta}(z_0) \left\{ -(a\eta^2)^{\eta\gamma} \cot(\pi\eta\gamma)
 \frac{\Gamma\big(1+\eta(1-\gamma)\big)}{\Gamma(\eta\gamma)} \right.
 \label{besselMellinPlus} \\ \nonumber
 & \hspace{3em} \left. + \frac{\esp{-\gamma t_0}}{\sin(\pi\eta)}
   \sum_{k=1}^\infty\frac{(-z_0^2/4)^k}{(k-1)!}
 \left[\frac{1}{(k-\eta\gamma) \, \Gamma(k-\eta)}
  - \frac{(z_0^2/4)^\eta \cos(\pi\eta)}{[k+\eta(1-\gamma)] \,
 \Gamma(k+\eta)} \right] \right\} \;.
\end{align}
\end{subequations}

\begin{figure}[!t]
  \centering
  \includegraphics[width=0.4\textwidth]{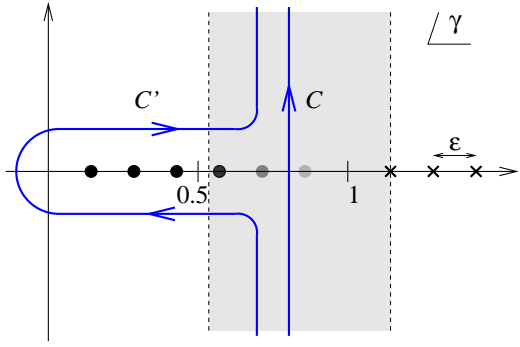}
  \caption{\sl Singularity structure of the Mellin transform
    $\Fm_\e(\gamma,t_0)$ in the complex $\gamma$-plane.  The shadowed region
    corresponds to the convergence strip of the Mellin transform; The crosses indicate
    the position of the singularities; the circles show the location of the
    poles of the terms in $\Fm_{\e}^{(+)}$; also shown are the original
    integration path $\contour$ in eq.~(\ref{intRep}), and the deformed contour
    $\contour'$ used in eq.~(\ref{osIntRep}).}
  \label{f:poles}
\end{figure}

I will show now that, with a proper choice of the contour $\contour$, only the
first term of $\Fm^{(+)}$ in eq.~(\ref{besselMellinPlus}) contributes to the
inverse Mellin transform~(\ref{intRep}) for $t > t_0$ --- the relevant region
for the on-shell limit.  Notice that the analytic continuation of $\Fm$ defines
a meromorphic function whose singularities are just the simple poles of
$\Fm^{(-)}$ at $\gamma = 1 + k\e : k = 1,2,\cdots$, as shown in
fig.~\ref{f:poles}. Actually, $\Fm^{(+)}$ is holomorphic in the whole plane
$\gamma \in \C$, since the poles at $\gamma = k\e : k=1,2,\cdots$ stemming from
the ratio $\cot(\pi\eta\gamma)/\Gamma(\eta\gamma)$ in the first line of
eq.~(\ref{besselMellinPlus}) are exactly canceled by those in the sum on the
second line; furthermore, the poles at $\gamma = 1 + k\e : k = 1,2,\cdots$
stemming from $\Gamma\big(1+\eta(1-\gamma)\big)$ in the first line are also
canceled by those in the sum on the second line.

It is convenient to compute the inverse Mellin transform separately for the
$(-)$ and $(+)$ pieces.  In the integral of
$\esp{\gamma t}\Fm^{(-)}(\gamma,t_0)$ one can close the contour path to the
left ($t-t_0 > 0$), without crossing any singularity, thus obtaining a vanishing
contribution, as expected.  Considering now the integral of
$\esp{\gamma t} \Fm^{(+)}(\gamma,t_0)$, one is not allowed to close
the contour either to the left or to the right, because the factor
$\esp{\gamma(t-t_0)}$ in front of the sum grows for $\Re(\gamma) \to +\infty$,
while the ratio of gamma-functions in the first term grows with $|\gamma|$ for
$\Re(\gamma) < 1/2 + \e/2$.  However, by folding the contour
$\contour\to\contour'$ so as to let it cross the real axis at some value
$\gamma_0 < \e$ (remember that $\Fm^{(+)}$ has no singularity), and then
computing the two contributions of eq.~(\ref{besselMellinPlus}) separately, one
obtains a vanishing integral from the second line, because the contour can be
closed to the left without crossing any singularities.

To summarize, with an integration contour $\contour'$ crossing the real axis at
$\gamma_0 < \e$ and going to infinity with $\Re(\gamma) > 1/2 + \e/2$ as in
fig.~\ref{f:poles}, only the first term in eq.~(\ref{besselMellinPlus})
contributes in the $\gamma$-representation~(\ref{intRep}) for $t > t_0$.

By performing the on-shell limit I end up with
\begin{align}
 \F_{\e}(t) &\equiv \lim_{t_0\to-\infty} \F_{\e}(t,t_0)
 \nonumber \\
 &=-\frac{\pi}{\Gamma(1+\eta)} \int_{\contour'} \frac{\dif \gamma}{2\pi\ui} \;
 \esp{\gamma t} (a\eta^2)^{\eta\gamma} \cot(\pi\eta\gamma)
 \frac{\Gamma\big(1+\eta(1-\gamma)\big)}{\Gamma(\eta\gamma)}
 \equiv  \int_{\contour'} \frac{\dif \gamma}{2\pi\ui} \; \esp{\gamma t}
 \Fm_\e(\gamma) \;,
 \label{osIntRep}
\end{align}
which is just the Mellin-Barnes representation~\cite{AS}(9.1.26) of the Bessel
function in eq.~(\ref{osBesselSol}).

{\bf 2)} It is straightforward to check that the on-shell Mellin transform
$\Fm_{\e}(\gamma)$ in eq.~(\ref{osIntRep}) obeys the homogeneous difference
equation
\begin{equation}\label{homFinDifEq}
 \Fm_{\e}(\gamma+\e) = a \chi(\gamma) \Fm_{\e}(\gamma)
\end{equation}
analogous to eq.~\cite{dimensional}(2.11). With some more effort, one can show
that the off-shell expression~(\ref{besselMellin}) obeys the inhomogeneous
difference equation~(\ref{finDifEq}).

{\bf 3)} The third issue concerns the validity of eq.~(\ref{marcelloSol}). By
explicitly computing the integral and the derivatives of $L(\gamma)$ in the
collinear model
\begin{align}
  \int_0^\gamma L(\gamma')\dif\gamma' &= \gamma\log(a) -\gamma\log(\gamma)
  +(1-\gamma)\log(1-\gamma) + 2\gamma \nonumber \\
  L^{(m)}(\gamma) &= (m-1)! [ (-1)^m \gamma^{-m} + (1-\gamma)^{-m} ]
  \qquad ( m \geq 1) \;,
\end{align}
the exponent within curly brackets in eq.~(\ref{marcelloSol}) becomes
($B_{2m+1} = 0 : m \geq 1$)
\begin{align}\label{collModAction}
  S(\gamma) &= \eta[\gamma\log(a) + 2\gamma -\gamma\log(\gamma)
  +(1-\gamma)\log(1-\gamma)] \\ \nonumber
  &\quad -\frac12\log(a) + \frac12\log(\gamma)  + \frac12\log(1-\gamma)
  + \sum_{m=1}^\infty \frac{B_{2m}}{2m(2m-1)} \left\{
  \left[\eta(1-\gamma)\right]^{1-2m} - \left[\eta\gamma\right]^{1-2m}
  \right\} \;.
\end{align}
The sum in the above equation is typical of the asymptotic expansion of the
logarithm of the gamma-function~\cite{AS}(6.1.40). In fact, by
comparing eq.~(\ref{collModAction}) with the asymptotic expansion
\begin{align}
  &\log\Gamma\big(\eta(1-\gamma)\big) -\log\Gamma(\eta\gamma) \approx
  \eta\left[ (1-2\gamma)\log(\eta) -1 +2\gamma -\gamma\log(\gamma)
  +(1-\gamma)\log(1-\gamma) \right] \nonumber \\
   &\qquad\qquad+ \frac12\log(\gamma) -\frac12\log(1-\gamma) + \sum_{m=1}^\infty
  \frac{B_{2m}}{2m(2m-1)} \left\{\left[\eta(1-\gamma)\right]^{1-2m}
  - \left[\eta\gamma\right]^{1-2m} \right\}
 \label{logGammaAsy}
\end{align}
one gets
\begin{equation}\label{expAction}
  \exp\{S(\gamma)\} \approx \frac{\esp{\eta}}{\sqrt{a}\,\eta^{\eta+1}}
  \; (a\eta^2)^{\eta\gamma} \;
  \frac{\Gamma\big(1+\eta(1-\gamma)\big)}{\Gamma(\eta\gamma)} \;.
\end{equation}
Apart from the irrelevant normalization factor
$\esp{\eta}/\sqrt{a}\eta^{\eta+1}$, eq.~(\ref{expAction}) agrees with the
integrand in eq.~(\ref{osIntRep}), when one takes into account that the
$\eta\to+\infty$ asymptotic expansion in powers of $1/\eta$ of
$\cot(\pi\eta\gamma)$ is a numeric constant ($\mp\ui$ according to the sign of
$\Im\gamma$).

\begin{figure}[!t]
  \centering\hfill
  \includegraphics[width=0.45\textwidth]{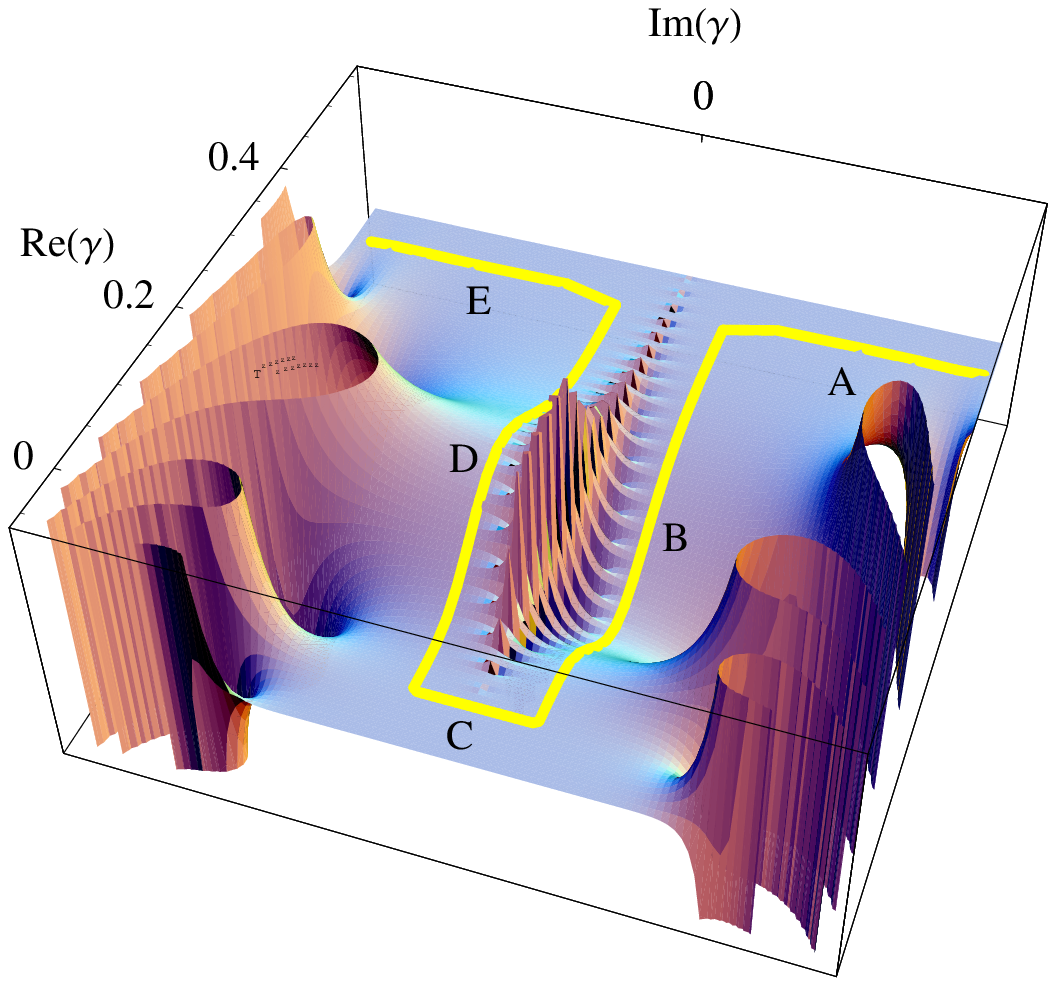}\hfill
  \includegraphics[width=0.45\textwidth]{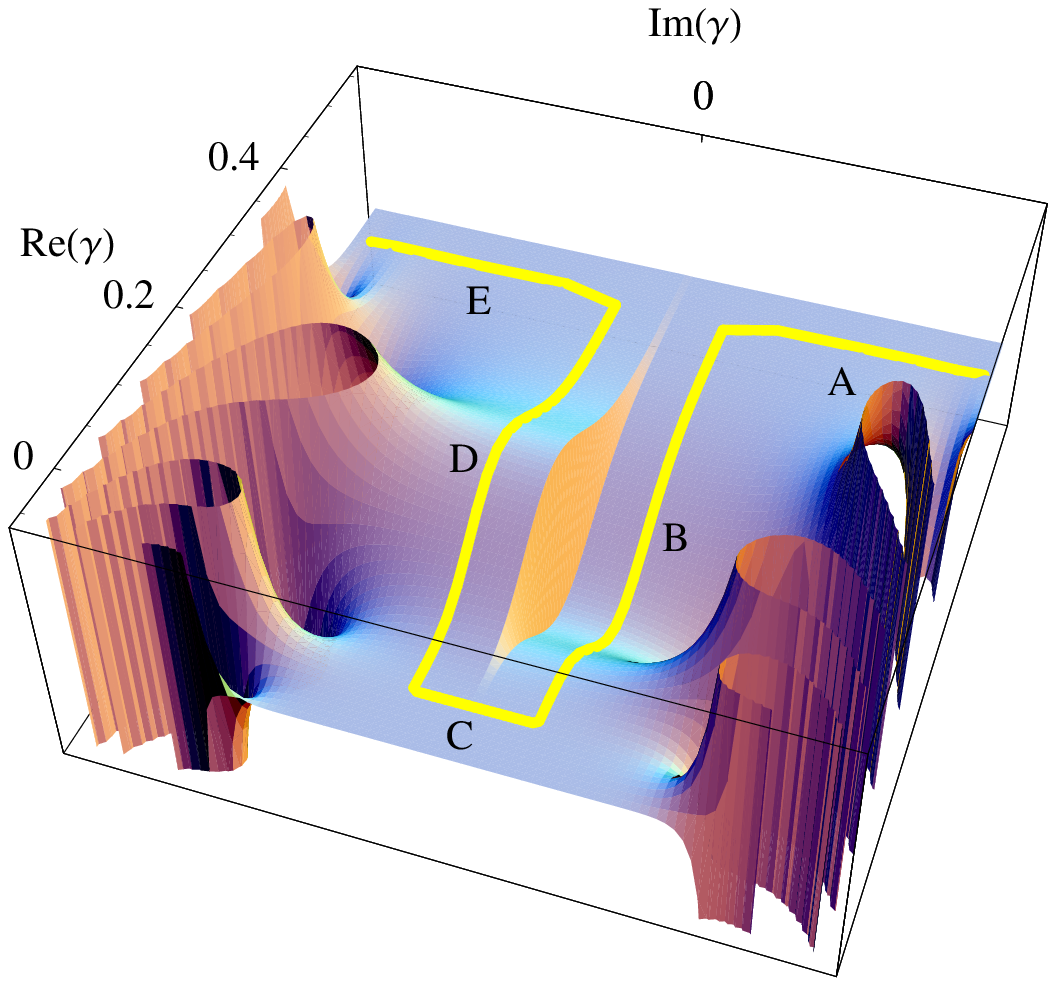}\hfill\null
  \caption{\sl a) The imaginary part of the integrand in eq.~(\ref{osIntRep})
    showing the singularities on the positive real semi-axis; in yellow a sketch
    of the fastest convergence path. b) asymptotic limit of the integrand
    showing the discontinuity~(\ref{realInt}) on the real axis with a peak
    around the saddle point value~(\ref{saddle}).}
  \label{f:path}
\end{figure}

{\bf 4)}
The last step is to evaluate the integral
in eq.~(\ref{osIntRep}) in the large-$\eta$ limit.  It turns out that, for small
values of $\e$ and values of $a\esp{\e t} < 1/\chi(1/2) = 1/4$, the fastest
convergence contour path surrounds the interval $0 < \Re(\gamma) < 1/2$
(cf.~fig.~\ref{f:path}) at a distance decreasing with $\e$. The main
contribution to the integral comes just from this region (parts B and D in
fig.~\ref{f:path}).
In the limit of vanishing $\e$, the string of poles at $\gamma = k\e$
accumulates into a branch-cut at $\gamma \in ]0,+\infty[$. In fact, while the
ratio of gamma-functions is regular at $\gamma > 0$ also in the
$\eta \to +\infty$ limit, the cotangent
$\cot(\pi\eta\gamma) \to -\ui\sign(\Im\gamma)$ becomes discontinuous across the
real axis with a jump equal to $-2\ui$.

Therefore, neglecting the contributions to the integral in eq.~(\ref{osIntRep})
from the parts A, C and E of the contour path, the contributions of B and D
amount to the integral in $\gamma\in ]0,1/2[$ of the discontinuity of the
integrand, which can be easily obtained by replacing $\cot(\pi\eta\gamma)$ with
$-2\ui$. One obtains
\begin{equation}\label{realInt}
  \F_{\e}(t) \approx \frac{1}{\Gamma(\eta+1)} \int_0^{1/2} \dif\gamma \;
  \esp{\gamma t} (a\eta^2)^{\eta\gamma}
  \frac{\Gamma\big(1+\eta(1-\gamma)\big)}{\Gamma(\eta\gamma)}
  \approx \frac{\sqrt{a}\,\esp{-\eta}\eta^{\eta+1}}{\Gamma(\eta+1)}
  \int_0^{1/2} \dif\gamma \; \esp{\gamma t + S(\gamma)} \;,
\end{equation}
where use have been made of eq.~(\ref{expAction}).

Some remarks are in order.
Firstly, by expanding in $\e\to0$ the prefactor in the
r.h.s.\ of eq.~(\ref{realInt})
$\sqrt{a}\,\esp{-\eta}\eta^{\eta+1}/\Gamma(\eta+1)=\sqrt{a/2\pi\e}[1+\ord{\e}]$,
eq.~(\ref{marcelloSol}) is correctly reproduced.
Secondly, the integrand, being a discontinuity of a solution of the difference
equation~(\ref{homFinDifEq}), is
itself a solution of the same equation.
Thirdly, the integral representation~(\ref{realInt}) of the on-shell density
$\F$ uses an integration path lying on the real axis.

The last remark explains the possibility of having a stable saddle point in the
real direction, despite the fact that our original integral in
eq.~(\ref{intRep}) involves a real analytic integrand and an integration contour
$\contour$ parallel to the imaginary axis.  In fact, according to the analysis
of~\cite{dimensional}, in the $\eta\to\infty$ limit the leading part of the
exponent in eq.~(\ref{realInt}) is given by
\begin{equation}\label{espLeading}
 \gamma t + S(\gamma) \simeq \gamma t + \eta \int_0^\gamma L(\gamma') \;
 \dif\gamma' + \ord{\eta^0} \;,
\end{equation}
having considered $t$ a possibly large parameter. The saddle point condition is
(cf.~eqs.~(\ref{sxEq},\ref{gb}))
\begin{equation}\label{saddle}
  \begin{cases}
    \e t + \log\left[a \chi(\gb)\right] & = 0\\
    \chi'(\gb) & < 0
  \end{cases}
  \quad \iff \quad
  \gb(a\esp{\e t}) = \frac{1-\sqrt{1-4a\esp{\e t}}}{2}
  = a\esp{\e t} + \ord{(a\esp{\e t})^2}
\end{equation}
and is fulfilled when $4 a\esp{\e t} < 1$ so that $0 < \gb < 1/2$. The saddle
point behaviour of the discontinuity of the original integrand $\Fm(\gamma)$ is
apparent in fig.~\ref{f:path}. The final result is
\begin{equation}\label{spInt}
 \F_{\e}(t) \approx \frac1{\sqrt{-\chi'(\gb)}}
 \exp\left\{\gb\,t + \frac1{\e}\int_0^{\gb} \log[a\chi(\gamma')] \;
 \dif\gamma'\right\} = \frac{a\esp{\e t}}{[1-4a\esp{\e t}]^{1/4}}
 \exp\left\{\frac1{\e}\int_0^{a\esp{\e t}}
 \frac{\dif \aaa}{\aaa}\; \gb(\aaa) \right\} \;,
\end{equation}
and exhibits the factorization of the $1/\e$ collinear singularities, thus
allowing us to derive the relation between the $\MSbar$ and $Q_0$ gluons, as
explained in the previous section.

In conclusion, the analysis of the frozen-coupling collinear model provides
analytic expressions for the gluon densities and anomalous dimensions in both
$\MSbar$- and $Q_0$-schemes which agree with the results of
ref.~\cite{dimensional}, sec.~2. In particular, the explicit expression of the
Mellin transform $\Fm_\e(\gamma)$ offers a concrete test of the asymptotic series
representation~(\ref{marcelloSol}), and also allows us to understand the
relation between the original Mellin integral~(\ref{osIntRep}) and the real-axis
integral~(\ref{realInt}), the latter being the basic tool to prove (by
saddle-point estimate) the factorization of collinear
singularities~(\ref{spInt},\ref{gAsy}).

\section{Collinear model with running coupling ($\boldsymbol{b > 0}$)
  \label{s:bpos}}

In this section I shall extend the collinear model to the more realistic
situation of running coupling. I shall show that most of the analysis
preformed in sec.~\ref{s:b0} for the fixed coupling case can be carried out with
running coupling too, with analogous results.

\subsection{Asymptotic behaviour of the solutions\label{ss:abs}}

Like the $b=0$ case, the solution of eq.~(\ref{intEqRc}) obeys a
second order differential equation:
\begin{equation}\label{diffEqRc}
 (1+AB\esp{\e t})\F'' - (1+2\e+AB\esp{\e t})\F' + [\e(1+\e)+A\esp{\e t}]\F
 = -A\esp{\e t_0} \delta(t-t_0) \;.
\end{equation}
In order to characterize the IR and UV regular solutions of the homogeneous
equation, I first determine their large-$|t|$ behaviour. This can be
accomplished by rewriting eq.~(\ref{diffEqRc}) in Schr\"odinger-like form and
then using the WKB approximation. In detail, by letting
\begin{equation}\label{rescale}
 \F(t) \equiv \frac{\esp{(\frac12+\e)t}}{1+AB\esp{\e t}} h(t)\;,
\end{equation}
we obtain for $h$ the Schr\"odinger equation
\begin{equation}\label{schroedinger}
 h'' - V h = 0 \;, \qquad
 V(t) = \frac14 -\frac1{B} + \frac1{B(1+AB\esp{\e t})} \;.
\end{equation}
The WKB approximation to the solution of eq.~(\ref{schroedinger}),
written in terms of the wave-number $\kappa \equiv \sqrt{V}$, reads
\begin{equation}\label{wkb}
 h(t) \simeq \frac1{\sqrt{\kappa(t)}}
 \exp\left\{\pm\int\kappa(t) \; \dif t\right\} \;,
\end{equation}
and yields, when inserted into
eq.~(\ref{rescale}), the two possible asymptotic behaviours of $\F(t)$.

In the IR region ($t\to-\infty$) we have
\begin{equation}\label{IRlim}
 \lambda \approx \esp{(\frac12+\e)t} \;, \quad V \to \frac14 \;, \quad
 \F \sim \exp\left[\left(\frac12+\e\pm \frac12\right)t\right] \;,
\end{equation}
and one identifies the IR regular solution as the one which vanishes more
rapidly, i.e., $\F_I \sim \esp{(1+\e)t}$.

In the UV region ($t\to+\infty$) we have
\begin{equation}\label{UVlim}
 \lambda \sim \esp{t/2} \;, \quad V \to \frac14 -\frac1B \;, \quad
 \F \sim \exp\left[\frac12\left(1\pm\sqrt{1-\frac4B}\right)t\right]
\end{equation}
and the solutions can have an exponential or oscillatory behaviour according to
whether $B$ is greater than or less than 4. In the former case, one again
identifies the UV regular solution as the one which vanishes more rapidly, i.e.,
$\F_U \sim \exp\Big[\frac12\Big(1-\sqrt{1-\frac4B}\,\Big)t\Big]$.

\begin{figure}[!t]
  \centering
  \includegraphics[width=0.5\textwidth]{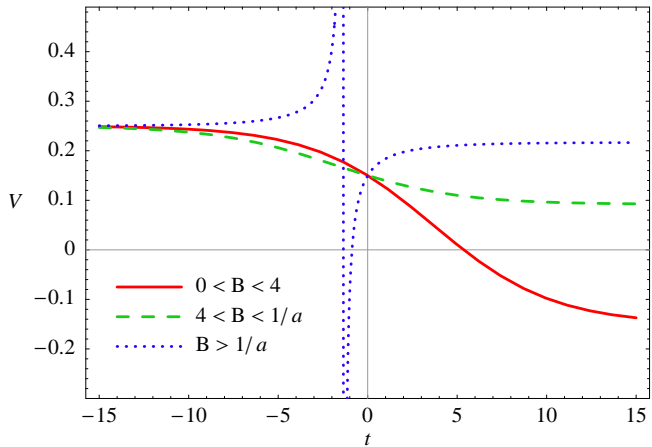}
  \caption{\sl The potential $V(t)$ of the Schr\"odinger
    equation~(\ref{schroedinger}) in the three running-coupling regimes: \regA\
    (solid-red), \regB\ (dashed-green) and \regC\ (dotted-blue).}
  \label{f:V}
\end{figure}

According to the value of $B$ --- and always considering $a<1/4$ --- it is
convenient to distinguish 3 regimes where the ``potential'' $V$ is qualitatively
different (cf.~fig.~\ref{f:V}):

\regA: $\qquad 0 < B < 4$\\
The potential is regular for all values of $t\in\R$, since $A>0$, and its UV
limit is negative:
\begin{equation} \label{Vinf}
 V(+\infty) = \frac14-\frac1B < 0 \;.
\end{equation}
As a consequence, in the UV region the wave-number $\kappa$ is pure imaginary
and all solutions share the same oscillatory behaviour, up to a relative phase.
The UV regular solution is thus undetermined.

\regB: $\qquad 4 < B < 1/a$\\
The potential is regular for all values of $t\in\R$, since $A>0$, and its UV
limit $V(+\infty)$ is positive. Hence, the wave-number $\kappa$ is real
throughout the whole $t$ range and the UV regular solution is uniquely
determined, as discussed above.

\regC: $\qquad 1/a < B$\\
The potential is singular at the Landau point $t_\Lambda$ of eq.~(\ref{landau}),
since $A<0$ and the denominator of $V$ vanishes at $t=t_\Lambda$. This
singularity might prevent the existence of a global solution for the integral
equation~(\ref{intEqRc}).  Nevertheless, The UV regular solution of the
differential equation~(\ref{diffEqRc}) can be unambiguously identified.

In the next section I shall explicitly determine the solution of the integral
equation~(\ref{intEqRc}) in the intermediate regime \regB, leaving to subsequent
sections the analysis of the regime \regA\ relevant in the limiting case
$b\to 0$ and of the regime \regC\ where the physical situation $\e\to 0$ at
fixed $b$ is recovered.

\subsection{Solution in momentum space\label{ss:smsB}}

By introducing the new variables
\begin{equation}\label{hypVar}
 \z \equiv -AB\esp{\e t} = -\frac{a(t)B}{1-a(t)B} \;, \qquad
 \F(t,t_0) \equiv -\z \Fh(\z,\z_0)
\end{equation}
the integral equation~(\ref{intEqGen}) becomes
\begin{align}
 \Fh(\z,\z_0) &= \frac1{B(1-\z)} \left[ \Theta(\z_0-\z)
 + \Theta(\z-\z_0) \left(\frac{-\z}{-\z_0}\right)^\eta \right]
 \nonumber \\
 &\quad - \frac{\eta}{B(1-\z)} \left[ \int_0^\z \Fh(\z',\z_0) \;\dif \z'
 + (-\z)^\eta \int_\z^{-\infty} (-\z')^{-\eta} \Fh(\z',\z_0) \;\dif\z'\right]\;.
 \label{hypIntEq}
\end{align}
By differentiating the above equation twice with respect to $\z$ yields the
differential equation
\begin{equation}\label{hypDiffEq}
 \z(1-\z)\Fh''
 + [(1-\eta)+(\eta-3)\z ] \Fh'
 - \left( \frac{\eta^2}{B} - \eta + 1 \right) \Fh
 = \frac{\eta}{B} \delta(\z - \z_0) \;,
\end{equation}
whose homogeneous version is just the hypergeometric differential equation with
parameters $u,v;w$ given by
\begin{equation}\label{hypPar}
 u,v \equiv 1-\frac{\eta}{2}\left(1\pm\sqrt{1-\frac4{B}}\right) \;, \qquad
 w \equiv 1-\eta = u + v -1 \;.  
\end{equation}

Let me now consider the regime \regB\ in which $4 < B < 1/a$ so that $u<v$ are
both real, $A>0$ and $\z < 0$ is a decreasing function of $t$
(cf.~eq.~(\ref{hypVar})).  At variance with the $b=0$ case, both IR ($\z\to0^-$)
and UV ($\z\to-\infty$) regular solutions of the homogeneous differential
equation~(\ref{hypDiffEq}), are unambiguously identified, as pointed out in
sec.~\ref{ss:abs}.  Explicitly
\begin{align}
 \Fh_U(\z) &\equiv c_U (-\z)^{-v}\hyp{2-u}{v}{v-u+1}{\frac1{\z}} \;,
  \qquad c_U \equiv \frac{\Gamma(2-u)\Gamma(1-u)}{\Gamma(1-w)\Gamma(v-u+1)}
 \label{FhU} \qquad \\
 \Fh_I(\z) &\equiv (-\z)^{1-w}\hyp{2-u}{2-v}{2-w}{\z}
 \label{FhI} \\
 W[\Fh_U,\Fh_I] &= (w-1)(-\z)^{-w}(1-\z)^{-2}\;.
 \label{hypWro}
\end{align}
By repeating the steps outlined in sec.~\ref{ss:sms}, the conditions of
continuity of $\Fh$ at $\z = \z_0$ and discontinuity of the first derivative
$-N(\z_0) = \eta/[B\z_0(1-\z_0)]$ provide the solution of eqs.~(\ref{hypIntEq})
and (\ref{hypDiffEq}):
\begin{equation}\label{hypSol}
 \Fh(\z,\z_0) = \frac{1-\z_0}{B(-\z_0)^\eta} \left[
 \Fh_I(\z)\Fh_U(\z_0)\Theta(\z-\z_0) +  \Fh_U(\z)\Fh_I(\z_0)\Theta(\z_0-\z)
 \right] \;,
\end{equation}

The integrated gluon density defined in eq.~(\ref{d:gluon}) can be computed in
closed form, and for $t > t_0$ reads (app.~\ref{a:igrc})
\begin{equation}\label{hypGluon}
 g_{\e,b}(t,t_0) = c_U \frac{\eta(1-\z_0)}{B(1-v)(-\z_0)^{\eta}}\Fh_I(\z_0) \,
 (-\z)^{1-v}\hyp{2-u}{v-1}{v-u+1}{\frac1{\z}}  \qquad (t > t_0) \;.
\end{equation}
showing also in this case a factorized structure. It is interesting to note that
the dependence of the equations and their solutions on $t$ occurs only through
$\z$, and because of eq.~(\ref{hypVar}), only through the combination $a(t)B$.

The on-shell limit $\z_0\to0^-$ of both integrated and unintegrated densities
can be easily computed by noting that $\Fh_I(\z_0)/(-\z_0)^\eta\to 1$, whence
\begin{align}
 \F_{\e,b}(t) &= \frac{-\z}{B}\Fh_U(\z)
  \label{osHypSol} \\
 g_{\e,b}(t) &= c_U \frac{\eta}{B(1-v)} (-\z)^{1-v}
  \hyp{2-u}{v-1}{v-u+1}{\frac1{\z}} \;.
  \label{osHypGluon}
\end{align}

The comparison of the unintegrated gluon density $\F$ with the perturbative
solution can be obtained rewriting the UV regular solution $\Fh_U$ as the sum
of two hypergeometric functions with argument $\z$ by means of the inversion
formula~\cite{AS}(15.3.7)
\begin{equation}\label{invFormula}
 \Fh_U(\z) = \hyp{u}{v}{w}{\z} +
 \frac{\Gamma(2-u)\Gamma(1-u)\Gamma(w-1)}{\Gamma(v)\Gamma(v-1)\Gamma(1-w)}
 \Fh_I(\z)
\end{equation}
and then using their series representation~\cite{AS}(15.1.1). The first term
yields
\begin{align}
 \frac{-\z}{B}\hyp{u}{v}{w}{\z}
 &= A\esp{\e t} \left\{ 1+ \sum_{n=1}^\infty (A\esp{\e t})^n \prod_{k=1}^n
  \frac{(u+k-1)(v+k-1)(-B)}{(w+k-1) k} \right\}
 \nonumber \\
 &= A\esp{\e t} \left\{ 1 +
 \sum_{n=1}^\infty(A\esp{\e t})^n\prod_{k=1}^n \left[ \chi(k \e) -B \right]
 \right\} \;,
 \label{hypPertSol}
\end{align}
and provides the perturbative expansion in terms of the parameter $A=a/(1-aB)$
and of the ``effective'' eigenvalue function $\chi(\gamma)-B$ relative to the
kernel defined in eq.~(\ref{d:KcB}).  The additional contribution to $\F(t)$ due
to the second term in eq.~(\ref{invFormula}) is of order
$(-\z)\Fh_I \sim (-\z)^{2-w} \sim (a\esp{\e t})^{(\eta+1)}$ and cannot find
place in the iterative solution, because it does not belong to the domain of the
kernel $K$ (cf.~sec.~\ref{ss:oslpe}). However, in the $\e \to 0$ limit, the
perturbative solution~(\ref{hypPertSol}) agrees with the exact
one~(\ref{osHypSol}) to all orders.

As last remark, the domain of convergence of the series in
eq.~(\ref{hypPertSol}) is finite ($\esp{\e t} < |AB|^{-1} = |1 - \e/b\asb|$),
unlike the $b=0$ case.

\begin{figure}[!t]
  \centering \hfill
  \includegraphics[width=0.47\textwidth]{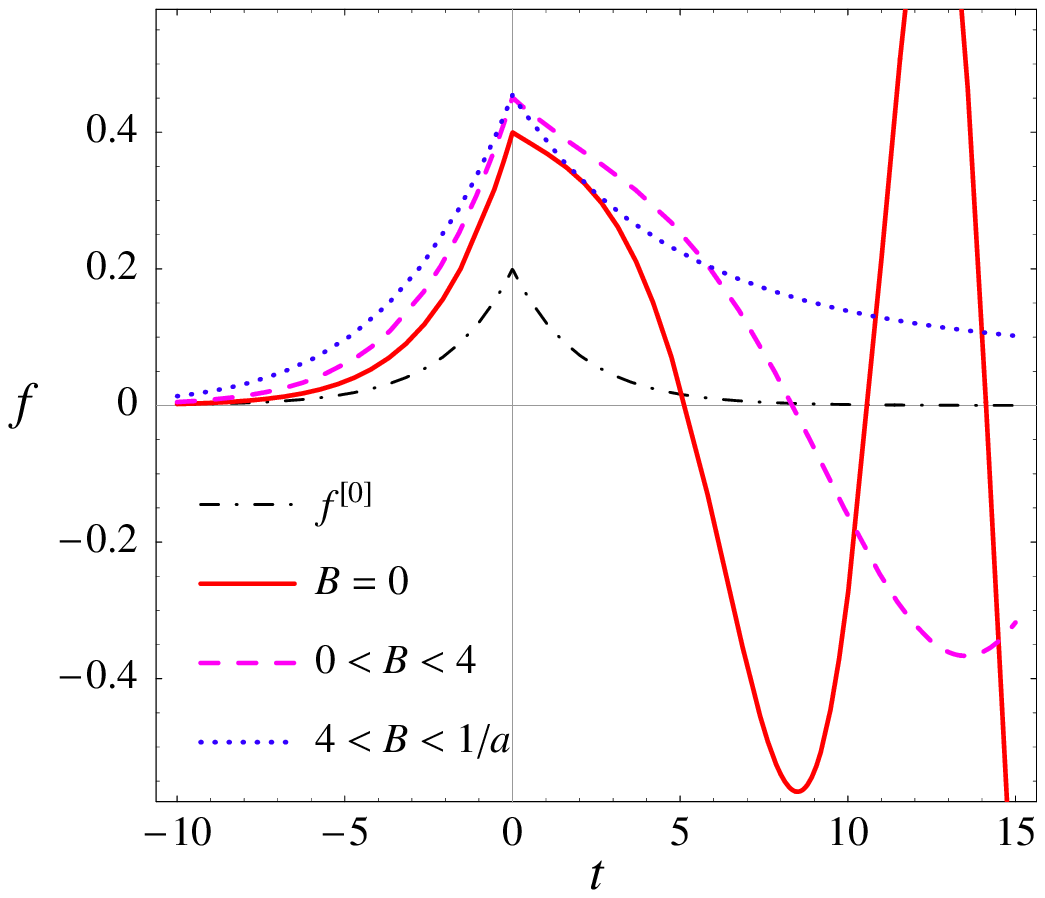} \hfill
  \includegraphics[width=0.47\textwidth]{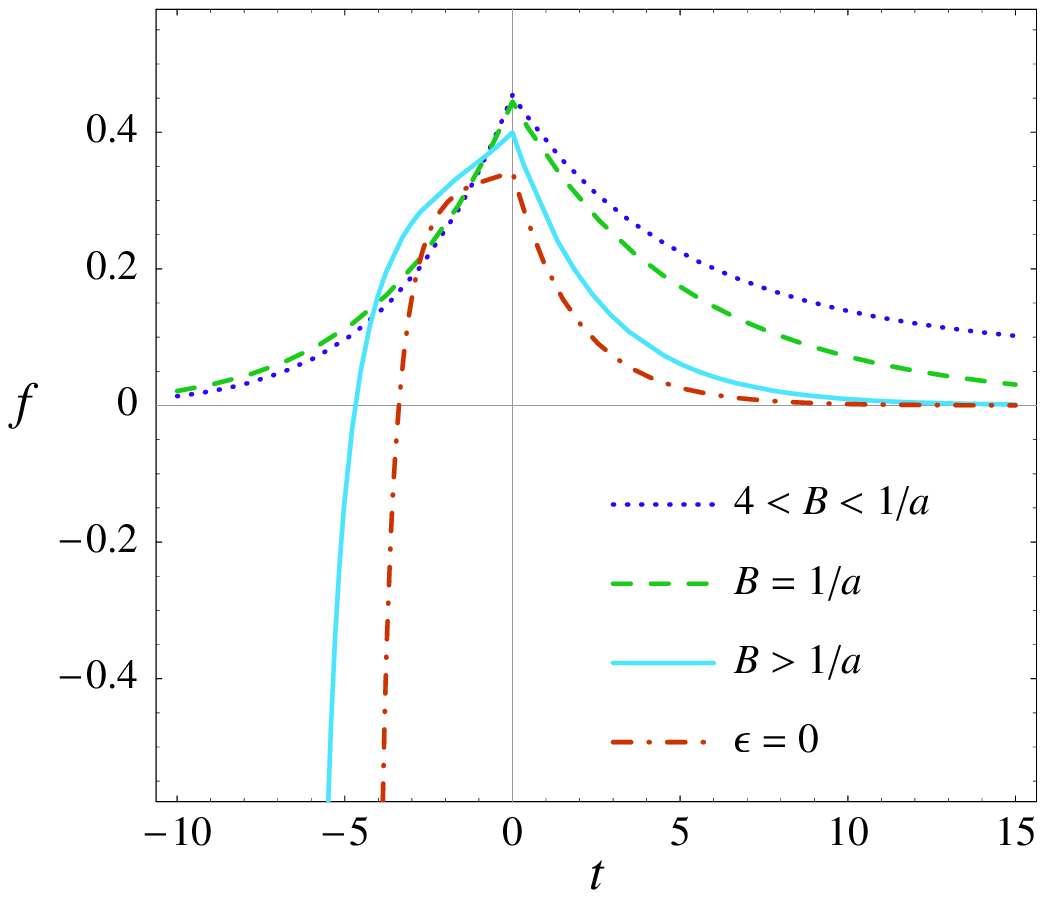} \hfill\null
  \caption{\sl Unintegrated gluon densities (times $\esp{-t/2}$) in various
    running-coupling regimes, with parameters $a=0.2$, $\eta=4.5$ and $t_0=0$.
    On the left side: lowest perturbative order (dash-dotted-black), fixed
    coupling (solid-red), regime \regA\ (dashed-purple), regime \regB\
    (dotted-blue); on the right side: again regime \regB\ (dotted-blue),
    boundary between \regB\ and \regC\ (dashed-green), regime \regC\
    (solid-cyan) and $\e \to 0$ limit at fixed $b = 1.2$ (dash-dotted-brown).
    The last two curves diverge at their Landau points.}
  \label{f:f}
\end{figure}

\subsection{Fixed coupling limit $\boldsymbol{b \to 0}$\label{ss:fcl}}

It is important to study the limit $b\to 0$ because, as already mentioned in
sec.~\ref{ss:sms}, it is not possible to determine the UV regular function in
the frozen coupling case. Actually, we saw in sec.~\ref{ss:abs} that this
problem is also present at $b > 0$ when $B < 4$ (regime \regA). Therefore, I
shall first derive the expression of the UV regular solution $\Fh_U$ for $B<4$
and then compute its limiting result at $B = 0$.

Clearly, $B = 4$ is a point of non-analyticity for the coefficients $u,\,v$ and
therefore also for the functions $\Fh_I,\,\Fh_U$.  When $B < 4$, the
coefficients $u,v$ become complex conjugate. It is easily verified that the IR
regular solution $\Fh_I$ remains real.  On the other hand, the expression for
$\Fh_U$ in eq.~(\ref{FhU}) yields two different (complex) results depending on
the choice $\Im u = -\Im v$ either greater or less than zero --- corresponding
to an analytic continuation from $B > 4$ to $B < 4$ in the complex variable
$B_\C$, from above ($B_\C = B + \ui 0$) or from below ($B_\C = B - \ui 0$).

Note that $B = 4$ is by no means a critical value for the coefficients of the
differential equations~(\ref{diffEqRc},\ref{hypDiffEq}): it only separates the
regimes of positive and negative effective potential in the UV region. Since
nothing prevents the existence of a real solution, it seems reasonable to define
the UV regular solution at $B < 4$ by taking the average of the two analytic
continuations:
\begin{equation}\label{d:FhUreg1}
 \Fh_U(\z;B<4) \equiv \frac12\left[\Fh_U(\z;B+\ui 0) + \Fh_U(\z;B-\ui 0)\right]
 = \Re\Fh_U(\z;B\pm\ui 0) \;.
\end{equation}
Of course, the definition~(\ref{d:FhUreg1}) joins continuously with the original
definition~(\ref{FhU}) at $B = 4$.  An explicit expression of the UV regular
solution for all $B > 0$ can be obtained by applying the
prescription~(\ref{d:FhUreg1}) to eq.~(\ref{invFormula}):
\begin{equation}\label{FhUreg1}
 \Fh_U(\z) = \hyp{u}{v}{w}{\z} + \Re\left(
 \frac{\Gamma(2-u)\Gamma(1-u)}{\Gamma(v)\Gamma(v-1)} \right)
 \frac{\Gamma(w-1)}{\Gamma(1-w)} (-\z)^{1-w}\hyp{2-u}{2-v}{2-w}{\z}
\end{equation}

The $B \to 0$ limit, at fixed $a,\,\e,\,t$ is then performed by exploiting the
series representation for the hypergeometric functions and the Stirling
approximation for the ensuing gamma-functions:
\begin{subequations}\label{B0limit}
\begin{align}
 \hyp{u}{v}{w}{\z} &\approx \Gamma(1-\eta) \left(\frac{z}{2}\right)^\eta
  J_{-\eta}(z) \\
 \hyp{2-u}{2-v}{2-w}{\z} &\approx \Gamma(1+\eta)
  \left(\frac{z}{2}\right)^{-\eta} J_\eta(z) \\
 \Re\frac{\Gamma(2-u)\Gamma(1-u)}{\Gamma(v)\Gamma(v-1)} &\approx
  \Re \esp{\ui\pi\eta}\left(\frac{\eta^2}{B}\right)^\eta =
  \cos(\pi\eta)\left(\frac{\eta^2}{B}\right)^\eta \;.
 \label{cosTerm}
\end{align}
\end{subequations}
Note in particular the cosine term in eq.~(\ref{cosTerm}) stemming from the real
part of the complex exponential: it is exactly the ``relative weight'' between
$J_\eta$ and $J_{-\eta}$ needed to build up the Bessel function of the second
kind $Y_\eta$, according to eq.~(\ref{FbU}). In conclusion, the substitution of
the expressions~(\ref{B0limit}) into eq.~(\ref{hypSol}) yields the
fixed-coupling result~(\ref{besselSol}), when the proper normalization factors
between $\Fb(z)$ and $\Fh(\z)$ are taken into account.

\subsection{Integrated gluon densities\label{ss:igdB}}

In this section I show how to derive explicit expressions for the integrated
gluon densities and anomalous dimensions in the $\MSbar$-scheme and $Q_0$-scheme
in the running coupling case. In particular, in this model it is confirmed the
claim of ref.~\cite{dimensional} that the running coupling corrections to the
$\MSbar$ anomalous dimension $\gamma^{(\MSbar)}\big(a(t),b\big)$ are
provided to all orders by the $\e$-dependence of the eigenvalue function
$\chi(\gamma,\e)$, according to the relation
\begin{equation}\label{MSanomDim}
  1 = a(t) \chi\big(\gamma^{(\MSbar)}, \, a(t) b \om \big) \;,
\end{equation}
where $\e$ has been replaced by $a(t) b \om$.

\subsubsection{$\boldsymbol{\MSbar}$-scheme}

The $\MSbar$ gluon for $b \neq 0$ is defined in two steps.  First, at fixed $b$
and $\e$, one perturbatively ($\asb \to 0$) computes the integrated gluon in
dimensional regularization. This implies that the calculation is naturally
performed with $\asb < \e/b$, i.e., $a < 1/B$, which corresponds to the regimes
\regA\ and \regB.  Then, all ensuing IR singularities appearing as poles at
$\e=0$ are isolated and factorized into an IR-singular ``transition function'',
to be identified with the $\MSbar$ gluon density $g^{(\MSbar)}(t)$.

The important point characterizing the $\MSbar$-scheme is the factorization
of $\e$-poles in the form
\begin{equation}\label{MSgluonB}
 g^{(\MSbar)}(t) = \exp\left\{\int_{-\infty}^{t} \dif \tau \;
 \gamma^{(\MSbar)}\big(a(\tau),b\big) \right\} = \exp\left\{ \frac1{\e}
 \int_0^{a(t)} \frac{\dif \aaa}{\aaa(1-\aaa B)}\;
 \gamma^{(\MSbar)}(\aaa,b) \right\} \;,
\end{equation}
where the $\MSbar$ anomalous dimension function $\gamma^{(\MSbar)}$ is required
to be $\e$-independent. The first integral in eq.~(\ref{MSgluonB}), which is
singular for $\e \to 0$ because of its IR lower bound, defines the $\MSbar$
anomalous dimension. The second integral is obtained by changing integration
variable according to eq.~(\ref{rce}), and is more suitable for comparison with
perturbative calculations.  If eq.~(\ref{MSgluonB}) contains all IR
singularities, the integrated gluon density~(\ref{osHypGluon}) of the collinear
model can be decomposed in the product
\begin{equation}\label{gluonFact}
 g_{\e,b}(t) = R_{\e}\big(a(t),b\big) \exp\left\{ \frac1{\e} \int_0^{a(t)}
 \frac{\dif \aaa}{\aaa(1-\aaa B)}\;
 \gamma^{(\MSbar)}(\aaa,b) \right\} \;,
\end{equation}
where the coefficient function $R$ is regular at $\e = 0$.

The above expression suggests a method for extracting the $\MSbar$ anomalous
dimension. One observes that the integrand in the exponent is singular at
$B=1/\aaa$.  On the other hand, for $B \to 1/a(t)$, no singularity
occurs in the off-shell functions $\F_{\e,b}(t,t_0)$ and $g_{\e,b}(t,t_0)$.
This signals that such singularity in the on-shell
limit is connected with the infinite evolution of $\tau$ from $t$ to $t_0 =
-\infty$, and therefore it affects only the exponential factor, while no such
singularity is expected in the coefficient function $R$. Therefore, if we take
the logarithmic derivative of $g$ with respect to $a(t)$ and subsequently the
limit $B\to 1/a(t)$ from below, we obtain ($a_t \equiv a(t)$):
\begin{equation}\label{formulaccia}
 \lim_{B\to 1/a_t} (1-a_t B)
 \frac{\partial{\log g}}{\partial a_t}
 = \lim_{B\to 1/a_t} (1-a_t B) \left[ \frac{\partial_{a_t}R}{R}
 + \frac{\gamma^{(\MSbar)}(a_t,B\e/\om)}{\e a_t (1-a_t B)} \right]
 = \frac{\gamma^{(\MSbar)}(a_t,\e/a_t \om)}{\e a_t} \;.
\end{equation}
Since the limit can be computed at any $a(t)$ and $\e$, the above formula
enables us to deduce the full functional dependence of $\gamma^{(\MSbar)}$ on
both $a$ and $b$.  In this model, from eq.~(\ref{osHypGluon}) we get
\begin{equation}\label{dlogG}
 \frac{\partial\log g}{\partial a_t} = \frac{\partial \z}{\partial a_t}
 \,\frac{\partial \log g}{\partial \z}
 = \frac{-B}{(1-a_t B)^2} \left[ \frac{1-v}{\z} - \frac1{\z^2}
 \frac{(2-u)(v-1)}{v-u+1}
 \frac{\hyp{3-u}{v}{v-u+2}{\displaystyle{\frac1{\z}}}}
 {\hyp{2-u}{v-1}{v-u+1}{\displaystyle{\frac1{\z}}}}
 \right] \;,
\end{equation}
where eq.~(\ref{hypVar}) and the differentiation formula for hypergeometric
functions~\cite{AS}(15.2.1) have been used. By performing the limit $B\to1/a_t$,
which implies $1/\z \approx a_t B-1 \to 0$, the second term within square
brackets does not contribute, and we end up with the simple expression
\begin{equation}\label{MSbarAnomDim}
 \gamma^{(\MSbar)}(a_t,b) = \left. \e a_t \lim_{B \to 1/a_t} (1-a_t B)
 \frac{\partial\log g}{\partial a_t} \right|_{\e = a_t b \om}
 = \e(1-v)|_{B=1/a_t,\e = a_t b \om} = \frac{1-\sqrt{1-4a_t}}{2} \;.
\end{equation}
which coincides with its fixed coupling ($b=0$) counterpart computed in
eqs.~(\ref{gammaMS}) and (\ref{saddle}).  This is a non-trivial result, since it
shows that the $\e$-independent kernel~(\ref{Kcoll}) provides a $b$-independent
$\MSbar$ anomalous dimension, according to eq.~(\ref{MSanomDim}).

Actually, it is not difficult to extend the collinear model to $\e$-dependent
kernels and check eq.~(\ref{MSanomDim}) in situations where $\gamma^{(\MSbar)}$
is explicitly $b$-dependent. For instance, by considering an $\e$-dependent
kernel
\begin{equation}\label{epsKernel}
 \Kc(\tau,\e) = \Xi(\e)[\Theta(-\tau)\esp{\xi(\e) \tau} + \Theta(\tau)]
 \;, \quad \chi(\gamma,\e) = \Xi(\e)
 \left(\frac1{\gamma}+\frac1{\xi(\e)-\gamma}\right) \;,
\end{equation}
where $\Xi$ and $\xi$ are regular functions of $\e$, one obtains the same type
of differential equation and hypergeometric solutions. Analogous expressions
hold for the gluon density, with the replacements $A \to A\Xi$, $B \to B/\Xi$
and with new parameters
\begin{equation}\label{epsPar}
 u,v \equiv 1-\frac{\eta\xi}{2}\left(1\pm\sqrt{1-\frac{4\Xi}{B\xi}}\right) \;,
 \qquad w \equiv 1-\xi\eta \;.
\end{equation}
The $\MSbar$ anomalous dimension is then straightforwardly obtained
(cf.~eq.~(\ref{MSbarAnomDim})):
\begin{equation}\label{EpsMSbarAnomDim}
 \gamma^{(\MSbar)}(a_t,b) = \e(1-v)|_{B=1/a_t;\;\e = a_t b \om}
 = \xi(a_t b \om)\frac{1-\sqrt{1 - 4a_t\,\Xi(a_t b \om)/\xi(a_t b \om)}}{2} \;.
\end{equation}
On the other hand, the solution $\gb(a_t,\e)$ of the implicit equation
$1 = a_t\chi(\gb,\e)$ (satisfying the perturbative condition $\gb(0,\e) = 0$) is
given by
\begin{equation}\label{EpsSaddle}
 \gb(a_t,\e) = \xi(\e)\frac{1-\sqrt{1 - 4 a_t\,\Xi(\e) / \xi(\e)}}{2}
 = \e(1-v)|_{B = 1/a_t}
\end{equation}
and exactly reproduces the anomalous dimension in eq.~(\ref{EpsMSbarAnomDim})
when substituting $\e \to a_t b \om$.

\subsubsection{$\boldsymbol{Q_0}$-scheme}

The $Q_0$-scheme gluon is defined by the $\e \to 0$ limit at fixed $b$ of the
off-shell $b$-dependent integrated density~(\ref{hypGluon}), analogously to the
frozen-coupling definition in eq.~(\ref{gQ0}). In this limit, $B \to +\infty$
and we enter the regime \regC, in which the variable $\z$ is positive.
Therefore, we must extend the expression~(\ref{hypGluon}) of the integrated
density from \regB\ to \regC. At the separation point $B = 1/a$, however, the
running coupling assumes a constant value $a(t) = a$, whereas the parameter $A$
(eq.~(\ref{rc})) and the variable $\z$ diverge.  In order to avoid the singular
point $B = 1/a$, we can change regime by means of an analytic continuation to
complex $B$ of the equations (\ref{hypPar}-\ref{osHypGluon}) obtained in \regB.
According to whether \regC\ is reached from the upper or lower half of the
$B_\C = B\pm\ui 0$ complex plane, $(-\z)^p \to \z^p \esp{\pm\ui\pi p}$ acquires
a phase of different sign. This translates into a discontinuity of the analytic
continuation of $g$ at values of $B > 1/a$. For $t>t_0$ we obtain
\begin{align}\label{gluonReg3}
 g_{\e,b}(t,t_0) &= \frac{\eta^2}{B(1-v)(u-v)} \z^{1-v}
 \hyp{2-u}{v-1}{v-u+1}{\frac1{\z}} g_I(t_0) \;,
 \qquad (t > t_0 \;, \; B > 1/a )\\ 
 & g_I(t_0) = (1-\z_0)\left[\z_0^{v-2} \hyp{u}{2-v}{u-v+1}{\frac1{\z_0}} \right.
 \\ \nonumber
  & \left. \qquad\qquad\qquad - \esp{\pm\ui\pi(u-v)}
 \frac{\Gamma(2-u)\Gamma(1-u)\Gamma(u-v+1)}{\Gamma(2-v)\Gamma(1-v)\Gamma(v-u+1)}
 \z_0^{u-2} \hyp{2-u}{v}{v-u+1}{\frac1{\z_0}} \right] \;.
\end{align}
From this equation we learn that:
\begin{itemize}
\item the analytic continuation of the gluon density is still factorized in its
  $t$ and $t_0$ dependence with the same UV ($t$-dependent) factor as in \regB;
\item the discontinuity affects only the IR factor $g_I(t_0)$, because of the
  phase $\esp{\pm\ui\pi(u-v)}$ in the second term;
\item the Landau pole shows up in the well known branch point of the
  hypergeometric function at $\z = \esp{\e(t-t_\Lambda)} = 1$, in both UV
  and IR parts.
\end{itemize}

Our main goal, though, is to obtain the anomalous dimension, which is known to
be independent from the IR properties of the theory, provided the ``hard scale''
$t$ is large enough. In other words, the {\em effective anomalous dimension}
\begin{equation}\label{effAnomDim}
 \gamma_{\mathrm{eff}}(t,t_0) \equiv
 \frac{\dot{g}_{\e,b}(t,t_0)}{g_{\e,b}(t,t_0)}
 = \frac{1-v}{\eta} \frac{\hyp{2-u}{v}{v-u+1}{\displaystyle{\frac1{\z}}}}
 {\hyp{2-u}{v-1}{v-u+1}{\displaystyle{\frac1{\z}}}} \;, \qquad
 (\dot{g} \equiv \partial_t g)
\end{equation}
is expected to depend only on $a(t)$ for $t \gg t_0, t_\Lambda$, where
$t_\Lambda$ eventually represents a cutoff that regularizes the Landau pole
and gives mathematical meaning to the gluon density $g$. In this model, thanks
to the factorization property of $g$, the effective anomalous dimension is
independent of $t_0$, hence independent on the details of the analytic
continuations to \regC, and needs not a regulator of the Landau pole. As a
result, the $Q_0$-scheme anomalous dimension
\begin{equation}\label{gQ0B}
 \gamma^{(Q_0)}\big(a(t)\big) \equiv \lim_{t_0\to-\infty} \lim_{\e\to0}
 \frac{\dot{g}_b(t,t_0)}{g_b(t,t_0)} = \lim_{\e\to0} \gamma_{\mathrm{eff}}(t)
\end{equation}
is just the $\e\to0$ limit of the effective anomalous dimension.

In order to compute $\gamma^{(Q_0)}$, I rewrite the hypergeometric functions in
eq.~(\ref{effAnomDim}) as~\cite{AS}(15.3.5)
\begin{equation}\label{rewriteF}
 \hyp{2-u}{v-1+n}{v-u+1}{\frac1{\z}} =
 \left(1-\frac1{\z}\right)^{1-n-v} \hyp{v-1}{v-1+n}{v-u+1}{\frac1{1-\z}}
 \;, \quad (n=0,1) \;.
\end{equation}
In this form, only the third parameter and the argument of the hypergeometric
function diverge in the $\eta\to\infty$ limit.  Finally, by using the limit
representation~\cite{HTF} of the Tricomi confluent hypergeometric function $U$
\begin{equation}\label{limitRep}
 \lim_{\tt c\to\infty}\hyp{\tt a}{\tt b}{\tt c}{-\frac{\tt c}{\tt z}} =
 {\tt z}^{\tt a} \, U(\tt a, a-b+1, z) \;,
\end{equation}
and the four-dimensional running coupling
$a(t) = a/(1 + a b \om t) \equiv 1/b\om\ti$, I obtain
\begin{equation}\label{bgQ0}
 \gamma^{(Q_0)}\big(a(t),b\big) = a(t)
  \frac{U(-1/b\om,0,\ti)}{U(-1/b\om,1,\ti)} \;, \qquad
 \ti \equiv t+\frac1{a b \om}
\end{equation}
which is indeed a function of $a(t)$ and $b\om$ only. Let me stress that the
expression above resums the running coupling corrections of the $Q_0$-scheme
anomalous dimension to all orders in $b$.

The $Q_0$-anomalous dimension~(\ref{bgQ0}) can be directly obtained in $D=4$
dimensions as well. In fact, the differential equation~(\ref{diffEqRc}) at
$\e=0$, in the $\ti$ variable reduces to the Kummer's equation, whose
independent solutions are the confluent hypergeometric functions
$\hspace{0.5em}F\hspace{-1.15em}_1\hspace{0.65em}_1(1-1/b\om,2,\ti)$ and
$U(1-1/b\om,2,\ti) = U(-1/b\om,0,\ti)/\ti$. The UV asymptotic behaviour
identifies $U(\ti)$ as the UV regular solution, i.e., as the unintegrated gluon
density $\F = \dot g$ --- up to an IR-dependent normalization constant. In turn,
$U(1-1/b\om,2,\ti)$ is the derivative of $b\om U(-1/b\om,1,\ti)$, which
represent therefore the integrated gluon $g$ --- up to the same IR-dependent
normalization. Finally, the ratio $\dot g/g$ yields the $Q_0$-anomalous
dimension as in eq.~(\ref{bgQ0}).

An important check comes from the $\gamma$-representation~\cite{gammaRep}
\begin{equation}\label{gammaRep}
 \F_{\e,b}(t) \propto \int\frac{\dif\gamma}{2\pi\ui}\;
 \esp{\gamma t - \frac{X(\gamma)}{b\om}}
 = \int\frac{\dif\gamma}{2\pi\ui}\;\esp{\gamma t}
 \gamma^{-1/b\om} (1-\gamma)^{1/b\om} \;, \quad
 X(\gamma) \equiv \int \chi(\gamma) \;\dif\gamma
 = \log\frac{\gamma}{1-\gamma}
\end{equation}
yielding, in the collinear model, the confluent hypergeometric function
$U(1-1/b\om,2,\ti)$ --- apart from a $t$-independent factor --- as first
observed by M.~Taiuti in her degree thesis~\cite{martina}. From the
representation~(\ref{gammaRep}), by means of the saddle-point method, one can
obtain the running coupling corrections at any given order in $b\om$.  In the
$b\to0$ limit, the $Q_0$ anomalous dimension~(\ref{bgQ0}) reduces to the frozen
coupling value $\gb\big(a(t)\big)$ of eq.~(\ref{gammaQ0}), and coincides with
the $\MSbar$ anomalous dimension~(\ref{MSbarAnomDim}).  Starting from $\ord{b}$,
the two factorization schemes provide different results. In particular, this
collinear model predicts a $b$-independent $\MSbar$-scheme anomalous dimension,
whereas the $Q_0$-scheme contains non-vanishing corrections, which agree with
the $b$-expansion in $D=4$ dimensions.

\subsection{Solution in $\boldsymbol{\gamma}$ space\label{ss:sgsB}}

In this section I show that the $\gamma$-representation~(\ref{intRep}) is valid
also at $b \neq 0$ and similar conclusions as for the $b = 0$ case can be drawn
in the asymptotic small-$\e$ expansion of the gluon density.

\begin{figure}[!t]
  \centering
  \includegraphics[width=0.4\textwidth]{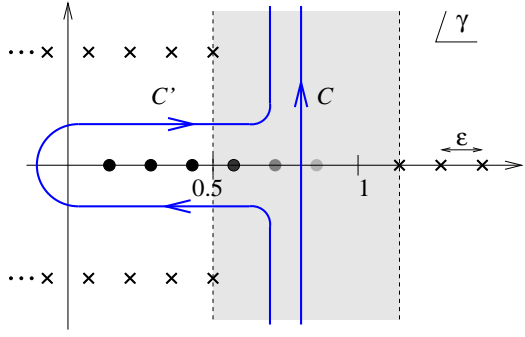}
  \caption{\sl Singularity structure of the Mellin transform
    $\Fm_\e(\gamma,t_0)$ in the complex $\gamma$-plane. Symbols as in
    fig.~\ref{f:poles}.}
  \label{f:hypPoles}
\end{figure}

The Mellin transform $\Fm_{\e,b}(\gamma,t_0)$ of the unintegrated gluon density,
defined in eq.~(\ref{d:Mellin}), exists in the strip
$1/2-\Re\sqrt{1/4-1/B} < \Re\gamma < 1+\e$ for all values of $\e$ and $B$.
However, the structure of the singularities of $\Fm(\gamma)$ is different from
that in eq.~(\ref{besselMellin}), since now $\Fm^{(+)}$ has two infinite series
of poles at $\gamma = (1\pm\sqrt{1-4/B})/2 - \e n : n \in \N$.  In particular,
when $B<4$ (regime \regA), these poles are located off the real axis, as shown
in fig.~\ref{f:hypPoles}.

Let me focus on the gluon density for $t > t_0$.  By following the same steps as
in sec.~\ref{ss:sgs}, the function $ \Fm^{(+)}(\gamma)$ can be decomposed in the
sum of a product of elementary and gamma functions plus hypergeometric functions
of type $\hypF{3}{2}$; by then deforming the integration contour $\contour$ into
$\contour'$ so as to cross the real axis at $\gamma_0 < \e$ (while leaving the
new complex poles to the left), the $\hypF{3}{2}$ functions do not contribute to
$\F_{\e,b}(t>t_0)$. In the on-shell limit $t_0 \to -\infty$ we are left with
\begin{align}\label{osHypIntRep}
 \F_{\e,b}(t) = \frac{\pi\eta}{B\Gamma(\eta)\Gamma(u)\Gamma(v)} \int_\contour
 \frac{\dif\gamma}{2\pi\ui} \;& \esp{\gamma t} (AB)^{\eta\gamma} \;
 \Gamma(u-1+\eta\gamma)\Gamma(v-1+\eta\gamma)
 \nonumber \\
 & \times \frac{\Gamma\big(1+\eta(1-\gamma)\big)}{\Gamma(\eta\gamma)}
 [ - \cot(\pi\eta\gamma) + \Re\cot(\pi u) ] \;,
\end{align}
where the real part in the last term descends from the real part used in
eq.~(\ref{FhUreg1}). With some trigonometric identities it is not difficult to
prove that, for $B < 4$,
\begin{equation}\label{ReCot}
 \Re\cot(\pi u) = \frac12[\cot(\pi u)+\cot(\pi v)] = \frac{
 \sin(\pi\eta)}{\cos(\pi\eta)-\cosh\big(\pi\eta\sqrt{\frac{4}{B}-1}\big)}
 \stackrel{\eta\to\infty}{\sim} \exp
 \left[-\pi\eta\textstyle{\sqrt{\frac{4}{B}-1}}\right] \;.
\end{equation}
When computing the inverse Mellin transform~(\ref{osHypIntRep}) along the
deformed path $\contour'$ in the large-$\eta$ (fixed $B$) limit, the term with
$-\cot(\pi\eta\gamma) \to \pm\ui\sign(\Im\gamma)$ becomes discontinuous on the
real axis and its main contribution is provided by the integral of such
discontinuity in the real interval $\gamma \in ]0, 1/2[$.  On the other hand,
the term proportional to $\Re\cot(\pi u)$ does not develop any discontinuity,
hence does not contribute in the parts B-C-D of the contour; furthermore in the
remaining parts A and E it is exponentially suppressed with respect to the other
term $\propto \cot(\pi\eta\gamma)$, as indicated in eq.~(\ref{ReCot}), and
therefore can be completely neglected. In conclusion
\begin{equation}\label{asyOsHypIntRep}
 \F_{\e,b}(t) \approx \frac{\eta}{B\Gamma(\eta)\Gamma(u)\Gamma(v)}
 \int_0^{\frac12}\dif\gamma \; \esp{\gamma t} (AB)^{\eta\gamma} \;
 \Gamma(u-1+\eta\gamma)\Gamma(v-1+\eta\gamma)
 \frac{\Gamma\big(1+\eta(1-\gamma)\big)}{\Gamma(\eta\gamma)} \;.  
\end{equation}

The main difference between eq.~(\ref{asyOsHypIntRep}) and its $b=0$ counterpart
eq.~(\ref{realInt}) is the presence of the two additional $u,v$-dependent
gamma-functions. The latter modify the analytic structure of the Mellin
transform $\Fm(\gamma)$ away from the real axis, but do not affect the mechanism
generating the discontinuity in the $\e\to 0$ limit. It is an easy exercise to
check that in the $b \neq 0$ case the integrand of eq.~(\ref{osHypIntRep})
(without $\esp{\gamma t}$) and its discontinuity~(\ref{asyOsHypIntRep}) obey the
homogeneous difference equation
\begin{equation}\label{BfinDifEq}
 \Fm_{\e,b}(\gamma+\e) = A \chi(\gamma,B) \Fm_{\e,b}(\gamma) \;.
\end{equation}

Finally, in the regime \regA\ we can verify the validity of the Laurent-series
representation~(\ref{marcelloSol}) also for the running coupling case. In fact,
by using in eq.~(\ref{asyOsHypIntRep}) the asymptotic expansion of the
gamma-functions in terms of Bernoulli numbers, after some calculation one indeed
reproduces eq.~(\ref{marcelloSol}) with
\begin{equation}\label{BrealInt}
 \Omega = \sqrt{\frac{A\eta}{2\pi}} \,
 \frac{\eta^{\eta+1/2}\esp{-\eta}}{\Gamma(\eta+1)} \,
 \frac{\Gamma(\eta/B)}{(\eta/B)^{\eta/b-1/2}\esp{-\eta/B}} \;, \quad
 L(\gamma,B) = \log[A\chi(\gamma,B)] \;.
\end{equation}
In the $\eta\to+\infty$ limit, $\Omega \to \sqrt{A/2\pi\e}$ and
\begin{equation}\label{BosLimit}
 \F_{\e,b}(t) \approx \frac1{\sqrt{2\pi\e}}\int\dif\gamma \; \esp{\gamma t}
 \frac1{\sqrt{\chi(\gamma,B)}}\exp\left\{\frac1{\e}\int_0^\gamma
 L(\gamma',B)\,\dif\gamma' +\frac{\e}{12}L'(\gamma,B) + \ord{\e^2}\right\}
\end{equation}
agrees with eq.~\cite{dimensional}(4.1).

To conclude this section, I have shown that the analysis of the collinear model
can be carried out explicitly in the presence of running coupling, and provides
analytic results for the anomalous dimensions in both $\MSbar$- and
$Q_0$-scheme, which agree with the general results of the literature and in
particular with the relations provided by secs.~3 and 4 of
ref.~\cite{dimensional}.
 
\section{Conclusions\label{s:c}}

In this article I have considered a simplified version of the integral equation
that determines the gluon Green's function in high-energy QCD in arbitrary
space-time dimensions $D=4+2\e$. The kernel of the integral equation agrees with
the true leading-$\log x$ BFKL kernel in the collinear limit, where the
transverse momenta of the gluons are strongly ordered. This model has no
phenomenological ambition, but embodies most of the qualitative features of the
real theory, e.g., the kinematical symmetry in the gluon exchange, the
leading-twist behaviour of the gluon density, the pattern of IR singularities
and the running coupling. It is therefore a useful tool to check and better
understand general results of the QCD literature. In fact, this model was
already considered in $D=4$ dimensions~\cite{collModel} for clarifying the
transition mechanism between the perturbative, non-Regge regime and the strong
coupling Pomeron behaviour.

In the present formulation I have explicitly determined the gluon densities and
their anomalous dimensions in two different factorization schemes: the
$\MSbar$-scheme, based on dimensional regularization, and the $Q_0$-scheme,
based on an initial off-shell gluon.  The main motivation for this analysis
stems from a previous work by M.Ciafaloni and myself~\cite{dimensional} where we
introduced a new method for solving the off-dimensional BFKL equation and for
performing the minimal subtraction of the collinear singularities. The rather
formal expressions we obtained and some sensible but unproven assumptions we
made, could be strongly supported by an explicit non-trivial example where they
are shown to be valid.

This analysis attains this object. In fact, in $4+2\e$ dimensions, the master
integral equation is solvable in terms of Bessel functions (with frozen
coupling) and hypergeometric functions (in the running coupling case). The
results obtained here are then often compared with series and integral
representations of ref.~\cite{dimensional}, showing their correctness and their
domain of validity. In particular, it is clarified the mechanism by which the
integral representation of the solution of the master equation --- a real
analytic function of the anomalous dimension variable $\gamma$ integrated along
a contour parallel to the imaginary axis --- is evaluated by a saddle point
integral along the real axis. I also show that the iterative/perturbative
solution to the master equation agrees with the exact one, at least up to the
order $n < 1/\e$; higher orders $n > 1/\e$ of the perturbative expansion do not
belong anymore to the domain of the kernel. Among the most important results
there is the confirmation of the formula~(\ref{MSanomDim}) determining the
$\MSbar$ anomalous dimension with running coupling from the $\e$-dependence of
the kernel, which in general was proved only up to $\ord{b^2}$
corrections~\cite{dimensional}.

On the whole, this model represents a useful tool for studying the mathematical
properties and the qualitative features of the off-dimensional BFKL equation,
even with running coupling. It supports the validity of the
procedure~\cite{dimensional} for determining anomalous dimensions in subleading
approximation, and encourages its application for extracting the leading-twist
anomalous dimension at full NL$x$ level.

\section*{Acknowledgments}

I am grateful to M.Ciafaloni for many interesting discussions and for his
encouragement during the preparation of this work.
This work has been supported by MIUR (Italy).

\appendix

\section{Integrated gluon density\label{a:igd}}

In this appendix I compute the off-shell integrated gluon density defined in
eq.~(\ref{d:gluon}), both at fixed coupling and with running coupling.

\subsection{Integrated gluon with frozen coupling\label{a:igdb0}}

At fixed coupling, it is convenient to use the $z$-variable introduced in
eq.~(\ref{besselVar}):
\begin{equation}\label{besselGluon}
 g_\e(t,t_0) = 1 + 2\eta\int_0^z x^{\eta+1} \Fb(x,z_0) \; \dif x \;,
\end{equation}
where the unintegrated density $\Fb$ is given in eq.~(\ref{besselSol}).
For $t < t_0$ we have~\cite{AS}(11.3.20)
\begin{align}
 g_\e(t<t_0) = 1+2\eta \frac{-\pi Y_\eta(z_0)}{4\eta\z_0^\eta}
 \int_0^z x^{\eta+1} J_\eta(x) = 1-\frac{\pi}{2}\left(\frac{z}{z_0}\right)^\eta
 J_{\eta+1}(z) Y_\eta(z_0) \;.
\end{align}
For $t > t_0$, the integral in eq.~(\ref{besselGluon}) is conveniently split
into 2 pieces
\begin{align}
 2\eta\int_0^z x^{\eta+1} \Fb(x,z_0) \; \dif x &=
  -\frac{\pi}{2z_0^\eta} \left[
  Y_\eta(z_0) \int_0^{z_0} x^{\eta+1} J_\eta(x) \; \dif x
  + J_\eta(z_0) \int_{z_0}^z x^{\eta+1} Y_\eta(x) \; \dif x \right] \\ \nonumber
 &= -\frac{\pi}{2z_0^\eta} \left[ z_0^{\eta+1}
  \left\{ Y_\eta(z_0) J_{\eta+1}(z_0) - J_\eta(z_0) Y_{\eta+1}(z_0) \right\}
  +J_\eta(z_0) z^{\eta+1} Y_{\eta+1}(z) \right] \;.
\end{align}
The terms in curly brackets are the opposite of the
Wronskian~(\ref{Wb}),~\cite{AS}(9.1.16)
\begin{equation}\label{pseudoWronskian}
 J_\eta Y_{\eta+1} - Y_\eta J_{\eta+1} = -W = \frac2{\pi z_0}
\end{equation}
and combined with the prefactors yield a $-1$ which cancels the $1+$ in the
definition~(\ref{besselGluon}) of the gluon. The final result reads
\begin{equation}\label{gluonBis}
 g_\e(t > t_0) = -\pi \frac{z}{2} \left(\frac{z}{z_0}\right)^\eta
 J_{\eta}(z_0) Y_{\eta+1}(z)
\end{equation}
and has the remarkable property of being factorized in the $z$- and
$z_0$-dependence.

\subsection{$\boldsymbol{\e \to 0}$ limit\label{a:igdb0e0}}

In the $\e \to 0$ limit of eq.~(\ref{gluonBis}), both the order and the argument
of the Bessel functions grow linearly with $\eta \to +\infty$. By writing the
argument in the form
\begin{equation}\label{besselArg}
 z = \eta s \;, \quad s \equiv 2\sqrt{a} \esp{\e t/2} \to 2\sqrt{a} \;,
\end{equation}
using the asymptotic expansions~\cite{AS}(9.3.6) of the Bessel function $J$ in
terms of Airy functions
\begin{align}
  J_\eta(\eta s) &\approx
  \left(\frac{4\zeta}{1-s^2}\right)^{\frac14}\eta^{-\frac13}
  \mathrm{Ai}(\eta^{\frac23}\zeta) \;, \qquad (\eta\to+\infty)
 \label{besselAiry} \\
 \frac23 \zeta^{\frac32} &\equiv I(s) \equiv
 \int_s^1 \frac{\sqrt{1-u^2}}{u}\;\dif u
 = \int_{\frac14}^{a\esp{\e t}}
 -\frac{\sqrt{1-4\aaa}}{2} \; \frac{\dif\aaa}{\aaa}
 = \log\frac{1+\sqrt{1-s^2}}{s}-\sqrt{1-s^2} \;,
 \label{Is}
\end{align}
and exploiting the asymptotic expansion~\cite{AS}(10.4.59) of the Airy function
\begin{equation}\label{airy}
 \mathrm{Ai}(x) \approx \frac1{2\sqrt{\pi}} \, x^{-\frac14}
 \exp\Big(-\frac23 x^{\frac32} \Big) \;, \qquad (x\to+\infty) \;,
\end{equation}
one obtains
\begin{equation}\label{Jasympt}
 J_{\eta}(z_0) \approx \frac{\exp\left[-\eta \; I(2\sqrt{a} \esp{\e t_0/2})
   \right]}{\sqrt{2\pi\eta}(1-4a)^{\frac14}} \;,
\end{equation}
where $I$ is the integral defined in eq.~(\ref{Is}).  In the same way,
by using the asymptotic expansion~\cite{AS}(9.3.6) of the Bessel function $Y$ in
terms of the Airy function $\mathrm{Bi}(\eta^{\frac23}\zeta)$ and the
large-$\eta$ expansion of the latter~\cite{AS}(10.4.63), one obtains
\begin{equation}\label{Yasympt}
 Y_\eta(\eta s) \approx
 -\frac{2\exp[\eta \;I(s)]}{\sqrt{2\pi\eta}(1-s^2)^{\frac14}} \;.
\end{equation}

Before applying the above formulas to the integrated gluon, we can immediately
derive the $\e\to0$ limit of the unintegrated gluon (cf.~eq(\ref{besselSol}))
for $t > t_0$. In fact
\begin{equation}\label{unintegrated}
 \F_\e(t > t_0) = -\frac{\pi}{\eta} \left(\frac{z}{2}\right)^2
 \left(\frac{z}{z_0}\right)^\eta Y_\eta(z) J_\eta(z_0)
 \approx \frac{a\;\esp{\frac{t-t_0}{2}}}{(1-4a)^\frac12} \exp\left\{\eta\left[
 I(2\sqrt{a} \esp{\e t/2}) - I(2\sqrt{a} \esp{\e t_0/2}) \right]\right\} \;.
\end{equation}
The difference of the integrals in the exponential yields
\begin{equation}\label{Idiff}
 I(2\sqrt{a} \esp{\e t/2}) - I(2\sqrt{a} \esp{\e t_0/2}) =
 -\frac1{\e} \int_{a\esp{\e t_0}}^{a\esp{\e t}}
 \frac{\sqrt{1-4\aaa}}{2} \; \frac{\dif\aaa}{\aaa}
 \quad \xrightarrow{\;\e\to0\;} \quad -\frac{\sqrt{1-4a}}{2}(t-t_0) \;,
\end{equation}
hence
\begin{equation}\label{unintegratedEps0}
 f(t>t_0) = \frac{a}{\sqrt{1-4a}} \exp\left[\frac{1-\sqrt{1-4a}}{2} (t-t_0)
 \right] = \frac{a}{\sqrt{1-4a}} \exp\left[\gb(a) (t-t_0) \right] \;,
\end{equation}
where $\gb(a)$ is the saddle-point value~(\ref{saddle}) at $\e = 0$.

As for the integrated gluon density $g_\e$, comparing eq.~(\ref{gluonBis}) with
eq.~(\ref{unintegrated}) we find
\begin{equation}\label{fgRatio}
 \frac{\F_\e(t > t_0)}{g_\e(t > t_0)}
 = \frac{z}{2\eta}\frac{Y_\eta(z)}{Y_{\eta+1}(z)} \;,
\end{equation}
which is nothing but the effective anomalous dimension in eq.~(\ref{gammaEps}).
We need the asymptotic behaviour of
\begin{equation}\label{modY}
 Y_{\eta+1}(\eta s) = Y_{\tilde\eta}\big((\tilde\eta-1)s\big)
 = Y_{\tilde\eta}(\tilde\eta \tilde{s}) \approx
 -\frac{2\exp[\tilde\eta\;I(\tilde{s})]}
  {\sqrt{2\pi\eta}(1-\tilde{s}^2)^{\frac14}} \;,
\end{equation}
where $\tilde\eta \equiv \eta+1$ and
$\tilde{s}\equiv s(1-1/\tilde\eta)\approx s(1-\e)$. From
\begin{equation}\label{tildeI}
 \tilde\eta\;I(\tilde{s}) \approx (\eta+1)\int_{s-\e s}^1 \frac{\sqrt{1-u^2}}{u}
 \; \dif u = (\eta+1)I(s) + \sqrt{1-s^2} = \eta I(s) +
 \log\frac{1+\sqrt{1-s^2}}{s}
\end{equation}
and $s\to 2\sqrt{a}$, we obtain a finite limit for the ratio~(\ref{fgRatio})
\begin{equation}\label{YYratio}
 \frac{\F_\e}{g_\e} \approx \sqrt{a} \exp[\eta\,I(s) - \tilde\eta\,I(\tilde{s})]
 \quad \xrightarrow{\;\e\to0\;} \quad
 \left.\frac{\sqrt{a}\;s}{1+\sqrt{1-s^2}}\right|_{s=2\sqrt{a}}\
 = \frac{1-\sqrt{1-4a}}{2} = \gb(a)
\end{equation}
which coincides with the saddle-point $\gb(a)$ at $\e = 0$. The explicit
expression for the off-shell integrated gluon density at $\e=0$ --- namely the
gluon in the $Q_0$-scheme --- is finally obtained dividing
eq.~(\ref{unintegratedEps0}) by $\gb(a)$, whence eq.~(\ref{gQ0}).

In the on-shell case, where the $t_0 \to -\infty$ limit is performed at
non-vanishing $\e$, the integrated gluon~(\ref{osGluon}) at large-$\eta$ behaves
like
\begin{align}
 g_\e(t) &\approx \frac{a}{\gb(a)(1-4a)^\frac14} \exp\left\{ \frac{t}{2} + \eta
 \left[ 1 + \frac{\log a}{2} + I(s) \right] \right\} \nonumber \\
 &= \ff(a) \exp\left\{ \eta\left[ 1+ 
 \int_1^{a\esp{\e t}} \frac{\dif \aaa}{2\aaa}
 - \int_{\frac14}^{a\esp{\e t}} \frac{\sqrt{1-4\aaa}}{2\aaa} \; \dif
 \aaa \right] \right\} \;.
 \label{osAsy}
\end{align}
where I used the Stirling approximation for gamma-functions and
eq.~(\ref{Yasympt}) in the asymptotic expansion, and an integral representation
for the exponent, together with the definition~(\ref{ffN}) for $\ff$, in the
last equality. It is possible to shift the lower limits of integrations to zero,
since the two logarithmic singularities at $\aaa = 0$ cancel in the sum of
the two integrals. The finite additional contribution is provided exactly by the
first term ``$1+$'' within square brackets. The final result is
\begin{equation}\label{collFact}
 g_\e(t) \approx \ff(a) \exp\left\{ \frac1{\e}
 \int_0^{a\esp{\e t}} \frac{1-\sqrt{1-4\aaa}}{2} \;
 \frac{ \dif \aaa}{\aaa} \right\} \;.
\end{equation}
Eq.~(\ref{collFact}) demonstrates the factorization of the collinear
singularities, and identifies the $\e$-finite coefficient factor $R\equiv\ff(a)$
and the $\MSbar$ gluon density, according to eq.~(\ref{gAsy}).

\subsection{Integrated gluon with running coupling\label{a:igrc}}

In the running coupling case, it is convenient to use the $\z$-variable
introduced in eq.~(\ref{hypVar}). The interesting kinematical region is at
$t > t_0$, which in the regimes \regA\ and \regB\ where $aB < 1$ corresponds to
$\z < \z_0 < 0$:
\begin{equation}\label{hyperGluon}
 g_{\e,b}(t>t_0) = 1-\eta\left[\int_0^{\z_0} \Fh(x,z_0) \; \dif x
 + \int_{\z_0}^\z \Fh(x,z_0) \; \dif x \right] \;,
\end{equation}
where the unintegrated density $\Fh$ is given in eq.~(\ref{hypSol}).

The first integral involves an integral of hypergeometric function of
type~\cite{AS}(15.2.4)
\begin{equation}\label{hypIntA}
 \int x^{c-2} \hyp{a}{b}{c-1}{x} \; \dif x = \frac{x^{c-1}}{c-1}\hyp{a}{b}{c}{x}
 \;, \qquad \begin{cases} a=2-u\\ b=2-v\\ c=3-w \end{cases} \;,
\end{equation}
where the condition of integrability at $x = 0$ is guaranteed by
$c-2 = \eta > 0$. The second integral in eq.~(\ref{hyperGluon}), after the
position $y=1/x$, involves an integral of type~\cite{AS}(15.2.3)
\begin{equation}\label{hypIntB}
 \int y^{a-1}\hyp{a+1}{b}{c}{y} = \frac{y^a}{a} \hyp{a}{b}{c}{y}
 \;, \qquad \begin{cases} a=v-1\\ b=2-u\\ c=v-u+1 \end{cases} \;.
\end{equation}
Summing the various contributions yields
\begin{align}
 g_{\e,b}(t>t_0) = 1 -&\frac{\eta(1-\z_0)}{B}
 \frac{\Gamma(2-u)\Gamma(1-u)}{\Gamma(1-w)\Gamma(v-u+1)}
 \label{bordel} \\
 &\times \left\{
 (-\z_0)^{1-v}\left[\frac1{w-2}\hyp{2-u}{2-v}{3-w}{\z_0}
 \hyp{2-u}{v}{v-u+1}{\frac1{\z_0}} \right. \right. \nonumber \\
 &\qquad\qquad\quad \left.\left.
 -\frac1{v-1}\hyp{2-u}{2-v}{2-w}{\z_0}\hyp{2-u}{v-1}{v-u+1}{\frac1{\z_0}}
 \right] \right. \nonumber \\ \nonumber
 & \qquad \left. + \frac1{v-1}\hyp{2-u}{2-v}{2-w}{\z_0} (-\z)^{1-v}
  \hyp{2-u}{v-1}{v-u+1}{\frac1{\z}} \right\} \;.
\end{align}
The hard task is to prove that the $\z$-independent terms, namely those stemming
from the square brackets in eq.~(\ref{bordel}), combine themselves in such a way
to give a $-1$ that cancels the $1$ at the beginning of the r.h.s..  The method
is to use relations between contiguous hypergeometric functions --- differing by
one unit in some of their $(a,b,c)$ parameters --- and their derivatives.

By introducing the short-hand notation
\begin{equation}\label{shortF}
 F_1 \equiv \hyp{2-u}{2-v}{2-w}{\z_0} \;, \qquad
 F_2 \equiv \hyp{2-u}{v}{v-u+1}{\frac1{\z_0}}
\end{equation}
and exploiting the relations~\cite{AS}(15.2.6) and~\cite{AS}(15.2.5), one gets
\begin{align}
 \hyp{2-u}{2-v}{3-w}{\z_0} &= \frac{2-w}{(1-u)(1-v)}[(1-\z_0)F_1'-F_1] \\
 \hyp{2-u}{v-1}{v-u+1}{\frac1{\z_0}}
 &= \left(1-\frac{2-u}{\z_0(1-u)}\right)F_2 + \frac{\z_0-1}{\z_0^2(1-u)}F_2' \;,
\end{align}
whence
\begin{equation}\label{quadrata}
 [\cdots]_{(\ref{bordel})} = \frac1{(1-u)(1-v)}\left\{
 (2-u)\left(1-\frac1{\z_0}\right) F_1 F_2 + (\z_0-1) \left(
 F_1' F_2 + \frac1{\z_0^2} F_1 F_2' \right) \right\} \;.
\end{equation}
In order to find a relation among the $F_j$'s and their derivatives, I exploit
the Wronskian~(\ref{hypWro}):
\begin{equation}\label{relFjFjp}
 \frac{W[\Fh_U,\Fh_I](\z_0)}{c_U} = (-\z_0)^{1-v-w} \left\{ \left(
 \frac{1+v-w}{\z_0} \right) F_1 F_2 + F_1' F_2 +\frac1{\z_0^2} F_1 F_2'
 \right\} \;.
\end{equation}
The combination $F_1' F_2 + F_1 F_2'/\z_0^2$ entering eq.~(\ref{quadrata})
can thus be expressed in terms of the product $F_1 F_2$. As a result, the terms
with $F_1 F_2$ cancel out and all gamma-functions simplifies.  Finally, by
substituting the explicit expressions~(\ref{hypPar}) of $u,v,w$, it is
straightforward to compute the sum of the $\z$-independent terms in
eq.~(\ref{bordel}) and to obtain $-1$, as I stated previously. The remaining
$\z$-dependent term provides the factorized expression~(\ref{hypGluon}) for the
integrated gluon density.


\end{document}